# Semi-automated image analysis of cellulose nanofibrils using machine learning segmentation and morphological thinning

Carlos Baez[1*], Udita Ringania[2], Saad Bhamla[2], Robert J. Moon[1*]

*1 The Forest Products Laboratory, USDA Forest Service, Madison, WI 53726, USA*

*2 Chemical and Biomolecular Engineering, Georgia Institute of Technology, Atlanta, GA 30332, USA*

*Carlos Baez ORCID iD: 0009-0008-3270-819X*

*Udita Ringania ORCID iD: 0009-0003-3543-8301*

*Saad Bhamla ORCID iD: 0000-0002-9788-9920*

*Robert J. Moon ORCID iD: 0000-0001-9526-0953*

*Corresponding author:*

carlos.l.baez@usda.gov

Robert.j.moon@usda.gov

## Abstract:

Reliable and rapid morphology mfeasurement of cellulose nanofibrils (CNFs) with a high level of branching and entanglement is crucial for quality control, grade definition, and investigating morphology-performance relationships in various applications. An image analysis framework, Fibril Analysis for Cellulose Technology (FACT), which utilizes machine learning (ML) segmentation and morphological thinning, was developed to measure the fibril width distribution of cellulose nanofibers (CNFs) from negative contrast scanning electron microscopy (NegC-SEM) images. The high-contrast and wide magnification range of NegC-SEM imaging enabled the capture of micro- and nanoscopic hierarchical branching structures of CNFs. Two ML approaches [Weka and U-Net] were used to create detailed binary segmentation of grayscale NegC-SEM images, critical for the width analysis. Morphological thinning was applied to the binary image to produce a 1-pixel-wide skeleton of the CNF fibril structure. Subsequently, the distance between the skeleton and the original fibril edge was used to calculate fibril width. The FACT framework was optimized and validated with idealized geometric and hierarchical branched structures. Additionally, two contrasting CNF morphologies (i.e., low and high branching) were analyzed. FACT effectively performed segmentation, skeletonization, and fibril width measurement of these CNF morphologies. FACT width results were comparable with manual measurements. Variations were attributed to differences in the number of width measurements per fibril. In the manual method, a single measurement is made per fibril. In contrast, FACT simultaneously makes multiple measurements along each fibril within the entire CNF branched network structure. The advantage of FACT is that complicated branching and network CNF structures can be measured without imparting any analyst bias in fibril selection and measurement. Additionally, once the ML model is trained, each image can be analyzed in under 5 minutes. The FACT code is publicly available in Zenodo.

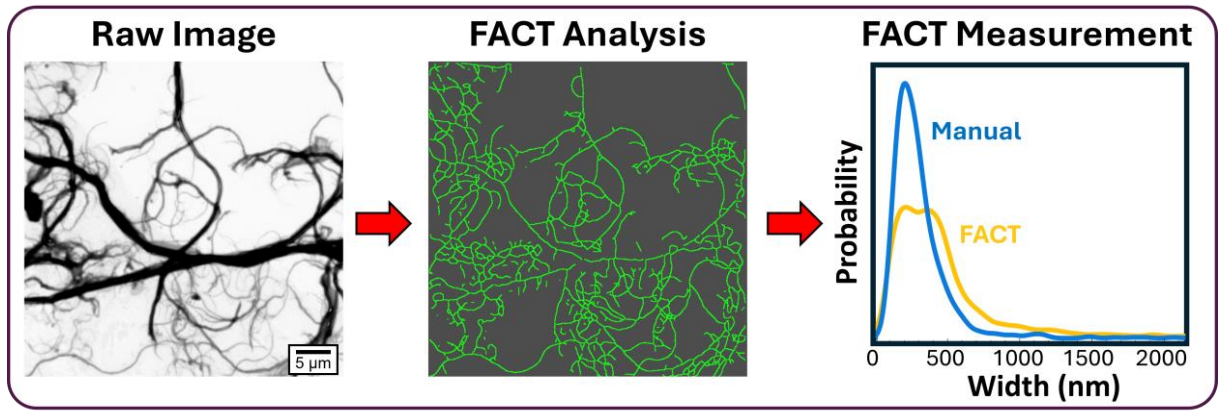

*Graphical Abstract:*

## Introduction:

Cellulose nanomaterials (CNMs) possess a unique set of characteristics that give them utility across a wide range of applications, including composite materials, biodegradable and renewable packaging, biomedical, and rheology modification (Moon et al. 2011; Zambrano et al. 2020; Chen et al. 2021; Li et al. 2021a; Li et al. 2021b; Eichhorn et al. 2022). Three predominant categories of woody plant-based CNMs are cellulose nanocrystals (CNCs), individual cellulose nanofibrils (iCNFs), and cellulose nanofibrils (CNFs) (Moon et al. 2023; Moon et al. 2025). CNCs primarily display a spindle-like, non-branching morphology, while iCNFs primarily display a high aspect ratio, non-branching fibril morphology. In contrast, CNFs are complex, fibril-like objects with extensive branching that entangle into networks.

Reliable and rapid measurement of CNF particle morphology (object size, degree of branching, and fibril width and length) is crucial for quality control, grade definition, and investigation of morphology-performance relationships for various applications. This need has been identified as a priority within the CNF research community (Moon et al. 2023). CNF suspensions exhibit a broad range of particle morphologies and dimensions, spanning from millimeters to nanometers. As a result, various analysis methods (direct imaging and indirect scattering approaches) are needed to quantify the morphology and dimensions across different length scales (Kangas et al. 2014; Moon et al. 2023). Macro- and micro-sized objects can be directly measured using fiber analyzers designed for pulp fiber measurement, which can typically detect object widths in the tens of microns and object lengths ranging from 100 µm to several millimeters. To directly image the nanoscale fibril features within CNF objects, techniques such as Transmission Electron Microscopy (TEM), Atomic Force Microscopy (AFM), or Scanning Electron Microscopy (SEM) are typically used. SEM has a much larger field of view and a broader magnification range. With SEM, it is possible to capture entire CNF objects while still providing sufficient resolution to capture the broad size-scale range of widths and lengths of hierarchical fibril branching within a single CNF object or across a tangled fibril network. SEM imaging offers the potential for enhanced characterization of CNF particle morphology (Moon et al. 2025).

However, using SEM to measure micro- to nanoscale features of CNFs remains challenging, as conventional imaging protocols require adjustments to be made. Firstly, improved sample preparation is needed to minimize agglomeration and overlap of branching fibrils, which negatively affect fibril identification and measurements during image analysis. Highly diluted suspensions can offset some of these problems (Ringania et al. 2022). Secondly, improved SEM imaging techniques are necessary for non-conducting objects to achieve higher image contrast and sharpness, thereby making fibril edges more distinct and facilitating the identification, measurement, and analysis of fibrils. Conventionally, non-conductive or beam-sensitive materials, such as CNFs, are imaged after they are coated with a thin conductive layer, typically 1-2 nm thick (Ang et al. 2020), which at higher imaging magnifications may obscure small features, fill gaps, artificially broadening objects, and typically results in lower contrast images. An alternative approach is to use the negative-contrast SEM (NegC-SEM) methodology, in which non-conductive objects are deposited onto an atomically smooth and highly conductive substrate (Mattos et al. 2019; Beaumont et al. 2021; Ringania et al. 2022; Moon et al. 2025). The secondary electron response for CNF is low, whereas it is high for the substrate, resulting in a high-contrast image that facilitates the analysis of CNF fibril length and width across various length scales. Thirdly, improved image analysis is necessary to measure a sufficient quantity of fibril features that are statistically relevant while minimizing analyst bias. This current study addresses these challenges by utilizing dilute suspensions, negative contrast scanning electron microscopy (Neg-C SEM), and developing a novel image analysis approach.

Current image analysis of CNFs often requires manual measurements and calculations, which are labor-intensive and time-consuming (Ang et al. 2020). There are various image analysis tools available, offering a wide range of capabilities for measuring particle dimensions. Manual image analysis with Gwyddion (Nečas and Klapetek 2012; Mattos et al. 2019) and ImageJ (Schneider et al. 2012; Ang et al. 2020) has been widely used for CNF image analysis. However, object selection and measurement are susceptible to user bias and fatigue, which contribute to the discrepancies in dimensional data of CNFs available in the literature. Semi-automated image analysis programs designed for the analysis of high aspect ratio fiber particles without branching morphologies, such as FibrilJ (Sokolov et al. 2017), FiberApp (Usov and Mezzenga 2015; Persson et al. 2017), DiameterJ (Hotaling et al. 2015), and a custom Python package (Willhammar et al. 2021), have had some success when applied to CNC or CNFs with minimal or low level of branching. However, when applied to CNFs with a high level of branching, these programs have struggled with the complexity of overlapping fibrils, varying degrees of fibrillation or branching, and fibrils of varying size scales. The MATLAB code UNDfiber, which was effective to some extent for highly dispersed optical images (Ringania et al. 2022), failed to capture any meaningful data when encountering overlapping fibrils in the SEM images. This current study addresses this gap by developing a semi-automatic image analysis approach called Fibril Analysis for Cellulose Technology (FACT), which can measure the width distribution of hierarchical, branched, and entangled CNF structures from NegC-SEM images.

This paper describes the development of FACT, which utilizes machine learning (ML) and morphological thinning tools to segment CNF fibril images and measure fibril width distributions. Additionally, it validates the width measurements using simulated fibril-like geometries. FACT was then used to analyze NegC-SEM images of CNFs from two published studies: one with a low level of branching (Mattos et al. 2019; Beaumont et al. 2021) and one with a high level of branching and networking (Ringania et al. 2022). FACT width results were compared with manual measurements. Where possible, this study followed the recommendations for sample preparation, imaging, analysis, and reporting as outlined in the paper "Perspectives on Cellulose Nanofibril Size Measurement Using Scanning Electron Microscopy" by an International Organization for Standardization (ISO) task group working on standards for CNF characterization (Moon et al. 2025).

## Materials and Methods

### Cellulose Nanofibrils (CNFs):

FACT was used to analyze NegC-SEM images of two variations of CNFs (low-level branching and high-level branching) from published studies. A brief description of the reported CNF material preparation from these studies is given here. The low-level branching CNFs used in the Mattos *et al.* (Mattos et al. 2019) and Beaumont *et al.* (Beaumont et al. 2021) studies were prepared from a never-dried, fully bleached, and fines-free sulfite birch pulp, without any chemical or enzymatic pretreatment, diluted to 0.5 wt.% solids in water, and fibrillated via six passes through a high-pressure microfluidizer (Microfluidics M110P). The resulting CNF suspension (0.5 wt.% solids in water) consists of a low-branched fibril structure, with fibril dimensions ranging between 5-30 nm in width and 50-5000 nm in length (Beaumont et al. 2021). The CNFs with high-level branching, as used in the Ringania *et al.* (Ringania et al. 2022) study, were produced by the Process Development Center at the University of Maine [lot # U-103, 90% fines]. The 'fines' percentage, defined by the ISO specification (ISO 2014), is the percentage of fibrils with a length less than 200 µm. These CNFs were prepared by mechanical fibrillation of wood pulp fibers (Masuko MKZB15-50J super mass colloider) without prior chemical or enzymatic pretreatment. The resulting CNF suspension (3 wt.% solids in water) consists of a hierarchical branched fibril structure, with fibril branching element dimensions ranging between 20-300 nm in width and overall branched particle size of several tens of micrometers (Ringania et al. 2022).

### Semi-Automated Image Analysis:

A semi-automated image analysis framework, Fibril Analysis for Cellulose Technology (FACT), was developed to measure the fibril width distributions for the entire CNF hierarchical branching structure. A flowchart of this process is illustrated in **Fig. 1.** The FACT framework consists of five parts: 1) image acquisition, 2) image segmentation, 3) morphological thinning, 4) skeleton refining, and 5) image analysis.

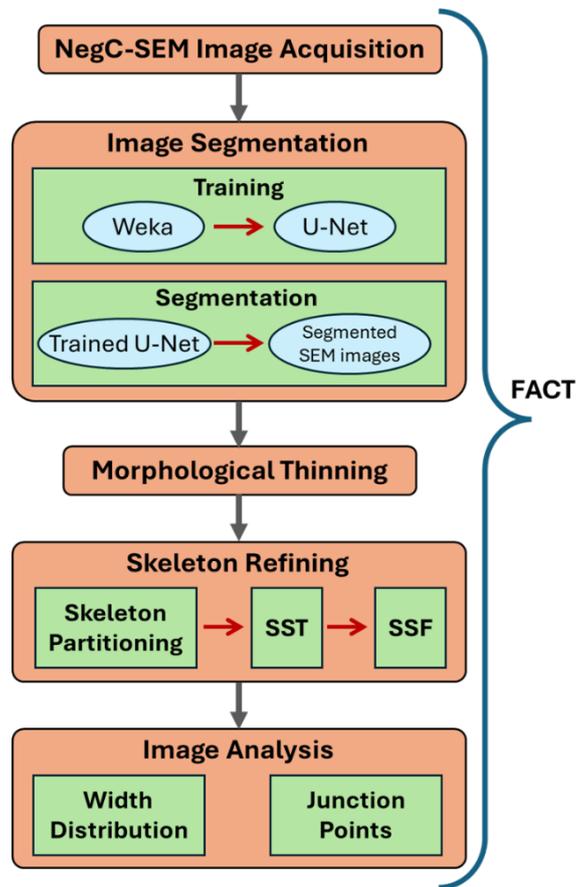

**Fig. 1** Flowchart of the FACT automated image analysis process. Where SST = skeleton segment trimming, and SSF = skeleton segment filtering

**NegC-SEM Image Acquisition:**

Image acquisition is critically important for the FACT approach to identify CNF branching and provide accurate width measurements of individual fibrils. Images should have high contrast, sharply defined CNF-substrate edges or boundaries, and a low area coverage (<20%) of well-dispersed CNFs on the substrate surface. NegC-SEM imaging of CNFs exhibits greater CNF-substrate contrast differences than conventional SEM imaging (Mattos et al. 2019; Beaumont et al. 2021; Ringania et al. 2022; Moon et al. 2025), which enables more accurate image segmentation.

In this study, FACT is used to analyze NegC-SEM images of two variations of CNFs (low-level branching and high-level branching) from published studies. A brief description of the reported SEM sample preparation and imaging parameters from these studies is given here. The low-level branching CNF samples used in the Mattos *et al.* (Mattos et al. 2019) and Beaumont *et al.* (Beaumont et al. 2021) studies were prepared by first diluting the starting 0.5 wt.% solid aqueous suspension to 0.01 wt.% using Milli-Q water and tip ultrasonicated (3 min with pulse on/off of 5/1 s at 10% amplitude). Electrically conductive substrates were produced using freshly cleaved mica discs that were spun-coated with a 4 nm thick layer of either gold, platinum/palladium alloy, or iridium. This substrate was then dipped in a 0.33% w/v Poly(ethylene imine) solution for 1 min, rinsed with Milli-Q water, and subsequently dipped in the CNF diluted suspension for 1 min. It was dried under ambient conditions (Mattos et al. 2019; Beaumont et al. 2021). NegC-SEM images were obtained using an SEM [FE-SEM, Zeiss

Sigma VP] at imaging conditions [1-1.5 kV, working distance of 1.8-6 mm] with an in-lens secondary electron detector. Images used in the current study (**Fig. 2a**) had either a 10,000 or 30,000 times magnification, with an image size of 2048 × 1347 pixels, corresponding to pixel resolutions of 5.43 nm/pix and 1.79 nm/pix, respectively.

High-level branching CNF samples used in the Ringania *et al.* study (Ringania et al. 2022) were prepared by first diluting the starting 3 wt.% solid aqueous suspension to 0.001 wt.% using DI water and subsequently mixed using a vortex mixer (VWR Analog Vortex Mixer No. 10153-838, speed 7, 30 s). A micro-pipette was used to deposit four 2 µL droplets of the diluted suspension onto a 1 cm x 1 cm silicon wafer (resistivity: 1-30 Ohm-cm, P-type with no $SiO_2$ top coating, 460-530 µm thickness, 2 nm roughness, ordered from Tedpella). These drops remained isolated even after mild spreading and then dried in ambient conditions overnight (for at least 12 h) before SEM imaging. NegC-SEM images were obtained using a benchtop SEM [Phenom Pure] at imaging conditions [5 kV, working distance of 8.8-8.9 mm] with a backscattering detector. Images used in the current study had a range of 1000 to 6000 times magnification, an image size of 2048 × 2048 pixels, and a pixel resolution ranging from 22 to 130 nm per pixel. The SEM's inbuilt automated image mapping (AIM) feature was used to capture the majority of a given droplet (**Fig. 2b**) and subsequently used to guide the location for higher magnification images, which were taken at various positions within individual droplets.

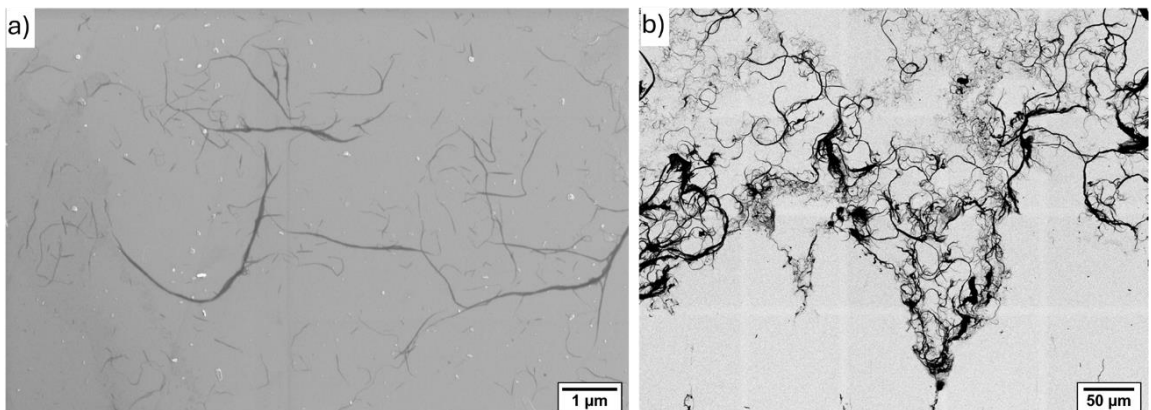

**Fig. 2** NegC-SEM images. **a** low-level branching CNFs by Beaumont *et al.* (Beaumont et al. 2021), and **b** automated image map of the high-level branching CNFs by Ringania *et al.* (Ringania et al. 2022), showing the notable differences in overall object scale, morphology, level of branching, and network structure

**Image Segmentation:**

The FACT image analysis program relies on binary segmented images, such as black-white contrast images that separate the foreground (e.g., CNFs) from the background (e.g., substrate) for subsequent measurements. The quality of the starting image, particularly the contrast difference between the foreground and background regions and the contrast variation within each image, dictates the approach and amount of effort necessary to produce segmented images that accurately represent the fibril structure and dimensions. No subsequent segmentation is needed for graphically simulated fibril structures, as the images are already binary. Traditional greyscale thresholding may be adequate for images of high contrast and low variation across the CNFs and the substrate. As the contrast decreases and/or variability increases across individual CNF particles, the CNF network, and the substrate, more sophisticated segmentation tools are required (e.g., Weka, U-Net).

For the grayscale NegC-SEM images used in the current study, low-level branching CNFs by Beaumont *et al.* (Beaumont et al. 2021), and high-level branching CNFs by Ringania *et al.* (Ringania et al. 2022), despite their high contrast, standard grayscale image thresholding resulted in suboptimal segmentation. Visual inspection found that fibrils were either not correctly identified or that substrate regions were identified as CNFs. Consequently, machine learning algorithms were employed to enhance the segmentation, utilizing the Weka segmentation plugin in ImageJ and the U-Net Convolutional Neural Network (CNN) implemented in Mathematica V12. Weka uses a library of machine learning algorithms to perform pixel classification. Weka is simple to use and fast to train, suitable for individual images. U-Net training was more involved and time-consuming, but the segmentation was more robust in terms of handling image-to-image contrast variability within and between the CNF and substrate regions. The trained U-Net model can be applied to multiple images (training and non-training images), significantly expanding image data sets that can be segmented, and its training can be continuously expanded upon.

For the low-level branching CNF material, five NegC-SEM images were received from the Beaumont *et al.* study (Beaumont et al. 2021). The Weka segmentation adequately segmented these images. Since this data set contained only five images and the Weka segmentation was adequate, a U-Net CNN was not trained on these images. The Weka plugin operates by having the user manually delineate regions as foreground or substrate using ImageJ drawing tools (**Fig. S1**), which are then used to train the Weka classifier. In this case, the two classes of interest are CNF and substrate. The Weka classifier then outputs a probability map (one image per class) where each pixel in each image has a value equal to the probability of belonging to a particular class. The class probability values are bound between zero and one. Pixel classification works by assigning each pixel a value representing a class that the user is interested in spatially identifying within an image (e.g., CNF or substrate). The segmented image is produced by thresholding one of the probability maps at the 0.5 level. If the quality of the final segmented image is sub-optimal, then Weka can be retrained by selecting additional features within the given image. This study found that using an individual model for each image yielded the best results.

For the high-level branching CNF material, 22 NegC-SEM images were received from the Ringania *et al.* (Ringania et al. 2022) study. It was found that the Weka segmentation was too noisy, which resulted in jagged edges around the finer fibrils. Consequently, a multi-step process was developed to improve the segmentation. First, Weka segmentation was completed on eight randomly selected grayscale NegC-SEM images to create binary images. Second, the noise in these binary images was minimized using a morphological operator to remove isolated foreground and background pixels. The resulting segmented images were thresholded at the 0.5 level to create "ground truth images" considered the "correct" segmentation. Third, the eight initial grayscale NegC-SEM images and their corresponding ground truth images were used to train a U-Net CNN implemented in Mathematica V12 to build a single machine-learning segmentation model. Finally, the trained U-Net model was used to segment all grayscale NegC-SEM images in this dataset.

The U-Net CNN used in this study was designed by Ronneberger *et al.* (Ronneberger et al. 2015) for pixel classification of biomedical images of HeLa cells. U-Net takes a single-channel grayscale image as input and outputs a multi-channel image, where each pixel contains the probability of belonging to a particular class (e.g., CNF or substrate). The original U-Net architecture was modified to reduce the total network size by decreasing the number of feature maps in each layer. This modification decreased the parameters from 31 million (124 MB) in the original U-Net to 7.7 million (31 MB) in our modified version. This modified network requires less GPU memory capacity for training and has the added benefit of reducing both network training and image segmentation time.

In practice, the network input image size is bounded by the available GPU memory. One approach to accommodate computation limitations is to decrease the network image input size by partitioning larger images into smaller sub-images. An overlap-tile strategy, as described by Ronneberger et al. (Ronneberger et al. 2015), was used to eliminate information loss when partitioning larger images. In the current study, the NegC-SEM images of the high-level branching CNF material had a size of 2048 × 2048 pixels. Because these images were too large, they were subdivided using the overlap-tile strategy into 700 × 700 pixel sub-images. For this, the image boundary was first padded by mirroring, resulting in a final image dimension of 2736 × 2736 pixels. Then, the padded image was partitioned into 64 sub-images of 700 × 700 pixels with an overlap of 342 pixels.

The U-Net CNN was trained using eight randomly selected initial greyscale NegC-SEM images and their corresponding segmented ground truth images (produced by Weka). One advantage of U-Net training is that very few acquired images are needed to yield precise segmentation. The current study partitions each image into 64 sub-images of 700 × 700 pixels. Additional training images were created by rotating each sub-image by 0, 90, 180, and 270 degrees and extracting the mirror image at the 0-degree orientation. These five operations increased the total number of training images from eight (2048 x 2048 pixels) to 2,560 (700 x 700 pixels) images, of which 62% were used for training, 33% for validation, and 5% for testing. U-Net was trained on these images using a stochastic gradient descent algorithm for 60 rounds. This training took approximately three hours on the computer used in this study (Intel Xeon Gold 5218R CPU running at 2.1GHz, 192 GB of RAM, and an NVIDIA RTX A5000 (24GB) graphics card).

After training, the U-Net segmentation of a given grey scale high-level branching CNF image consists of partitioning the original image to sub-images using the overlap-tile strategy (as described above), classifying every pixel into either CNF or substrate, and reassembly of the sub-images back together into a segmented image having the original starting dimensions, in this case, 2048 × 2048 pixels. The U-Net segmentation took approximately 5 s for each image.

**Morphological Thinning:**

Morphological thinning (Wolfram Research 2010a) is applied to skeletonize the binarized NegC-SEM images. The thinning operator is defined as the intersection between the original image and the complement of the hit-and-miss operator. The hit-and-miss operator (a Mathematica built-in function) performs a template-matching operation, utilizing selected structuring elements as input and rastering these elements throughout the image to look for a match (Wolfram Research 2008). The structuring element must correspond with the current image subsection's foreground and background pixels. If a match is found, the pixel beneath the origin of the structuring element is set to 1; otherwise, it is set to 0. Each iteration of the thinning operation effectively erodes the foreground objects in the binary image by one pixel. The thinning operator is iteratively applied until convergence, yielding the image skeleton.

Subsequently, the distance transform of the unfiltered binary SEM image is computed. In this process, the value of each foreground pixel is substituted by its Euclidean distance to the nearest background pixel. In cases where the particles in the binary image have jagged edges, it is recommended to apply a Gaussian blur filter to smooth the edges. FACT allows the user to select a kernel size for this filter. Recommended kernel size values are 0 (no filter) or 2. The thinning operator is applied to the filtered binary image. Then, the coordinates of the resulting skeleton pixels are used to extract their corresponding values from the distance transform image of the unfiltered binary.

**Skeleton Refining:**

The sensitivity of the thinning operator to sharpen foreground edges sometimes resulted in undesirable skeleton segments, which subsequently corrupt the quality of the CNF width information extracted from the image skeleton. A skeleton refinement method was developed that maintains the integrity of the original binary image while improving fibril width analysis. The skeleton is refined in three steps: 1) skeleton partitioning, 2) skeleton segment trimming (SST), and 3) skeleton segment filtering (SSF). Skeleton partitioning splits the interconnected skeleton into separate segments by identifying all the skeleton junction points using the hit-and-miss operator described earlier. This results in an image containing only the junction points. The pixels of the junction point image are then dilated from single pixel points into clusters of nine pixels using a dilation operator to form a 3x3 matrix with the original junction point in the center. Then, the skeleton image is subtracted by the edited junction point image, effectively partitioning the continuous skeleton network (**Fig. 3**). Finally, pixel connectivity-based methodologies (Wolfram Research 2010b) are used to extract data from each skeleton segment, such as pixel count, pixel positions, pixel values, and centroids.

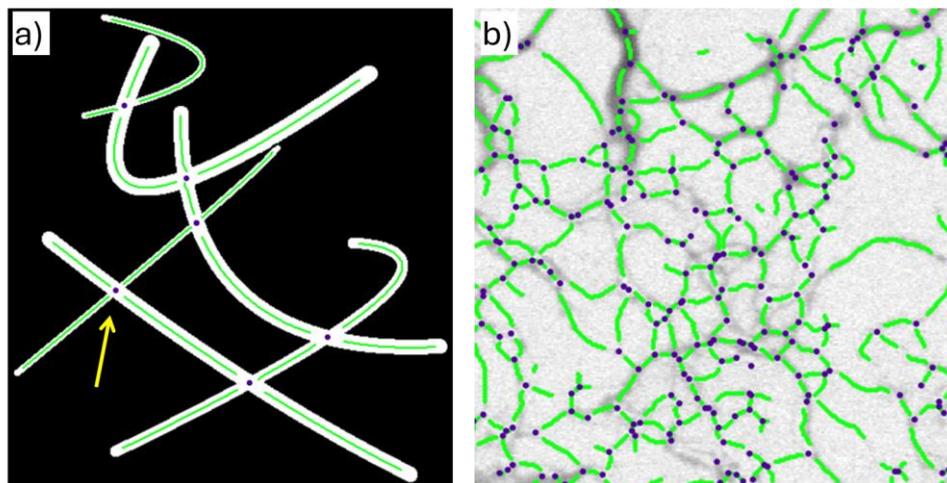

**Fig. 3** Examples of skeleton refinement of fibril networks analyzed by FACT, resulting in skeleton junction points (purple dots) and segments (green lines). **a** Simulated fibril structure (SST =20%, SSF = 1°), and **b** NegC-SEM image of highly branched CNFs (SST =25%, SSF = 25°). Skeleton segment trimming (SST) removes pixels from the ends of skeleton segments, thus increasing the gap between the green skeleton segments and the purple skeleton junction points (yellow arrow in part a)

The SST is used to eliminate pixels along the skeleton segment near the skeleton junction points, as they can incorrectly represent the width of the corresponding CNFs in this region. In SST, the points of a skeleton segment are first sequentially ordered, and then a fraction of the skeleton segment points are discarded from the skeleton segment ends, resulting in shorter, trimmed skeleton segments. This process is replicated using an identical skeleton segment endpoint discard fraction for all skeleton segments within an image. One aspect of this approach is that the number of pixels removed per segment is not constant but instead scales directly with the length of the skeleton segment. Thus, for a given SST value, the amount of pixel removal in the vicinity of junction points will vary. The selected SST fraction should balance maximizing fiber width counts while minimizing the number of skeleton counts within the junction region. The optimal SST fraction will vary for each image, depending on the nature of the network structure and branching (e.g., density of CNFs, etc.), as well as the fibril features the user wants to measure. SST value is determined by visual inspection and adjusted by trial and error. In the current study, the following SST values were used: 10%, 30% or 50% for simulated fibril structures, 10% for low-branched CNF images, and 25% for highly branched CNF images.

The SSF is used to eliminate unwanted small skeleton segments that do not track along the center line of a fibril and likely originate from segmentation edge defects (e.g., steps, discontinuities) along a given fibril (see segments labeled 2 and 3 in **Fig. 4**). For a given skeleton segment, its constituent points are plotted as {x, y} in two-dimensional (2D) space, where 'x' represents the skeleton segment position index, and 'y' is the corresponding distance transform value (DTV) at that position. Then, for each segment, its points are standardized to zero mean and unit variance, and the best-fit line and its slope ($\Phi$) are computed. This slope ($\Phi$) represents the average rate of change of the DTV along a specific skeleton segment. SSF uses the absolute value of this slope $|\Phi|$ to differentiate between wanted and unwanted skeleton segments. The value of SSF is determined by visual inspection and adjusted by trial and error. A balanced approach is needed when it comes to filtering skeleton segments. It is an imperfect process as some extraneous skeleton segments will inevitably remain, and if too much filtering is used, many correct segments could also be removed. There was no explicit cutoff $|\Phi|$ to separate wanted and unwanted skeleton segments. In most cases, a low $|\Phi|$ corresponds to skeleton segments that track along the length of individual fibrils, and the subsequent width measurements are representative of that fibril (label 1 in **Fig. 4**). In contrast, a high $|\Phi|$ corresponds to skeleton segments that do not track along the center line of a given fibril (label 2 and 3 in **Fig. 4**), and thus subsequent width measurements are not representative of any fibril. In the current study, the following SST values were used: 10 degrees for simulated fibril structures, 20 degrees for low-branched CNF images, and 25 degrees for highly branched CNF images.

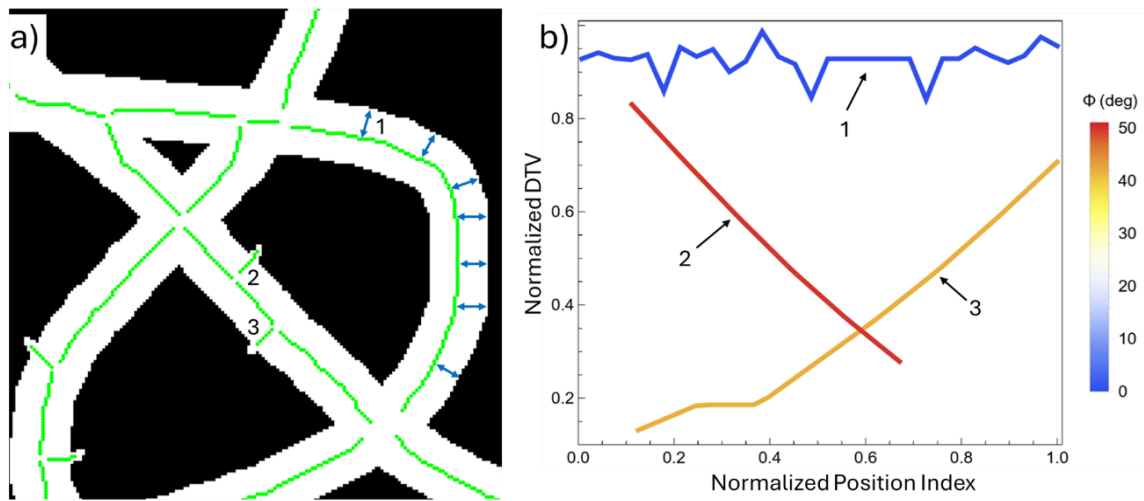

**Fig. 4** Skeleton segment filtering (SSF) process. **a** Simulated fibril structure showing how the skeleton segments follow the fibril profile. The blue double-ended arrows, labeled 1, represent the distance from the skeleton to the fiber edge. Simulated edge defects in the periphery of the fibril produce spurious skeleton segments, such as segments labeled 2 and 3. **b** Normalized distance transform value (DTV) versus normalized position index. The slope is color-coded.

The |Φ| is low for skeleton segments that follow the "fibril" centerline (labeled 1). In contrast, for the spurious segments (labeled 2 and 3), the |Φ| is high, and FACT removes these from the analysis. SST = 5%, and SSF refinement was not applied in this example, so that the unwanted skeleton segments would be shown in part a.

**Measurement & Analysis:**

FACT measures several features within each image: the pixel resolution (nm/pixel), the number of junction points, the number of skeleton segments, the number of pixels within skeleton segments, the Euclidean length of each skeleton segment, the width value for every pixel within all skeleton segments, and the average width for the pixels within each skeleton segment. In FACT, the width of a fibril is defined as twice the radial distance from the skeleton segment to the nearest background pixel (e.g., the outer edge of the fibril). Fibril width values are extracted from the distance transform image of the unfiltered binary using the skeleton coordinates. The limit of FACT width measurement resolution is 2 pixels, a consequence of defining width as doubling the radial distance from the skeleton segment to its outer edge. Therefore, a fibril must be two pixels wider than another fibril for FACT to detect a width difference between them. For this reason, the histograms presented in this report will have a bin width of 2 times the image pixel size (rounded to the nearest integer or multiple of 10). Higher-resolution imaging or imaging at higher magnifications can improve the number of pixels per fibril width. Digital zooming (post-processing) of captured images will not increase the number of pixels per fibril width.

Three approaches to analyzing FACT data are considered: skeleton segments, skeleton junction points, and individual pixels. However, relating these measurements to specific features that describe fibril and network structures can be problematic. The issue stems from the variability in skeleton junction formation, which is dependent on the quality of image segmentation, skeletonization, skeleton refinement, and the complexity of the fibril particle and network structures. In general, as the fraction of skeleton junction locations deviates from fibril branching or overlapping points, the less relevant the skeleton junction and segment analysis are in describing the fibril material analyzed. For simulated fibril structures, skeleton junctions are predominantly at branching and overlapping points, and minimal SST and SSF are needed. Thus, it is probable that skeleton junction analysis can be strongly related to branching and/or network density, while the skeleton segment analysis can be related to fibril length and width, branching length and width, network density, and each segment might represent a single fibril or branch and be used to count each feature.

In contrast, the grayscale NegC-SEM images of CNFs will have segmentation errors within regions of ambiguity, either between other fibrils or with the substrate, and the CNF particle and network can be very complex, requiring more aggressive SST and SSF. These errors significantly increase the number of skeleton junctions produced at locations unrelated to branching or crossover points. Additionally, fibrils will be sectioned into multiple skeleton segments, and thus, skeleton segments no longer represent the length or number of fibrils within the system.

Individual pixel analysis measures the width at every pixel within all the skeleton segments. The resulting data is not relatable to a specific fibril but to the "fibril collective" within the image. Every fibril will be measured multiple times along its length. However, the fraction of pixels measured will not be consistent across all fibrils. This inconsistency is a result of the number of junction points along the fibril length, fibril aspect ratio, morphology, SST, and SSF. Assuming all fibrils within a CNF image have a similar aspect ratio, FACT width statistical analysis will mostly depend on the total fibril length at each width level.

FACT can be applied to either individual images or multiple images as a batch. Batch image analysis is a sequential process that employs a consistent approach and parameters for segmentation, morphological thinning, and skeleton refinement. FACT outputs a statistical values table, a histogram, a box-whisker chart for each image within the batch, and skeleton overlay images (as displayed in this manuscript). Results are exported in Excel file format (*.xlsx) and as a tabulated/delimited file format (*.csv). The user can group the statistical results to compare the distribution of results between sample groups from the same batch of images. Additionally, users can modify the FACT code to measure features other than those listed above.

## Results and Discussion

FACT was developed to analyze CNF particles from NegC-SEM images and give a detailed assessment of the fibril width distributions of the hierarchical branching and entangled fibril structures typical of CNF materials. The FACT approach was systematically investigated to validate the effectiveness of fibril identification and width measurement of 1) simulated fibril structures, 2) micrographs of uniform diameter wires, 3) NegC-SEM images of low-level branching CNF material by Beaumont *et al.* (Beaumont et al. 2021), and 4) NegC-SEM images of high-level branching CNF material by Ringania *et al.* (Ringania et al. 2022). Additionally, the relevance of skeleton segment and junction analysis for describing fibril length, branching, and network density is discussed.

### Validation with Idealized Structures

The effectiveness of the FACT approach on width measurements was assessed and validated using idealized simulated fibril structures. Simulated fibril structures were used to assess the effects of object aspect ratio (**Fig. S2**), length (**Fig. S3**), shape (**Fig. S4**), orientation (**Fig. S5, S6**), junction points (**Fig. S7**), hierarchical level branching using rectangular segments (**Fig. 5**), and hierarchical level branching using curved segments (**Fig. 6**) on the FACT analysis. The geometry of an object influences the resulting skeletonization, which subsequently alters the number of pixels from which measurements are made, as well as the width measurements themselves. The effect of object aspect ratio on the skeleton counts results from the morphological thinning during skeletonization, in which pixel removal around the periphery of the object is limited by the smaller dimension as the periphery converges into a central line. In general, as an object's aspect ratio increases (**Fig. S2**), its skeleton segment length increases, which results in a greater fraction of pixel counts contributed by the object. However, for a given fixed aspect ratio, the number of skeleton pixel counts as a fraction of the object length is slightly higher for shorter objects as compared to longer objects (**Fig. S3**). In addition, objects with rounded ends will contribute a slightly higher fraction of pixel counts than objects with square ends of the same aspect ratio and length (**Fig. S4**). These effects are reduced when comparing fibril segments of higher aspect ratios (greater than 10). When analyzing CNF materials, the variations in aspect ratio, fibril length, and fibril end shape are not expected to have a significant influence, as fibril aspect ratios are commonly greater than 10, and ends are predominantly tapered.

Object orientation with respect to the image coordinate axis (e.g., pixel axis) can affect object skeleton pixel counts and, to a small extent, the width measurements. The number of pixels that comprise a given skeleton segment decreases with increasing object misalignment with the image coordinate axis and thus will be underrepresented in the final width measurement distribution of the analyzed structure. The minimum occurs at an off-axis angle of ±45 degrees, resulting in approximately 30% lower pixel counts. A correction factor was calculated by evaluating the change in skeleton pixel counts as a function of orientation and is given in **Fig. S5**. Additionally, there is an edge effect at the periphery of off-axis objects as pixels align stepwise, resulting in slight width variations along the object of approximately ±1.25 pixels (**Fig. S6**).

Skeleton junction points result from fibril branching, crossover or overlapping fibrils, and as a result of discontinuities along the fibril length (**Fig S7**). The total pixel count of skeleton segments can be minimally affected by the number of skeleton segments for low SST setting. For the idealized fibril shown in **Fig S7a**, the total number of pixels within the skeleton segment was 1962, while for a fibril of similar length but containing 4 junction points (**Fig S7b**) the total number of pixels within the 5 skeleton segments was 1950. A more consequential effect of junction points created from discontinuities is that they complicate associations to branching density, network density, and relating skeleton segments to fibril lengths or to fibril counts.

A four-level hierarchical branched structure was analyzed using FACT (**Fig. 5**). Each branch is a rectangular object with an aspect ratio of 10; all branching levels have the same total length, and the width of the four branch levels is constant at 30, 62, 120, and 240 pixels, respectively. FACT was able to measure the distinct widths for each branch level and adjust for orientation effects on the probability values **(Fig. 5b)**. The effect of branch orientation and connectivity (i.e., isolated versus connected objects) on skeleton counts is shown in **Fig. S8**. Each branch level is expected to contribute 25% of the total skeleton pixel counts to the width statistics (**Fig. S8a**). However, levels 2, 3, and 4 are oriented at ± 45 degrees and contribute fewer skeleton pixel counts than expected (**Fig. S8b**). To correct for this orientation effect, the average angle with respect to the vertical axis for each skeleton segment is calculated, and then the corrective factor is applied to resample the skeleton pixel counts (**Fig. S5**). In the case of isolated branches, the orientation correction adjusted the skeleton counts so that each level contributed approximately 25% (**Fig. S8b,e**). However, for connected branches (**Fig. 5b, S8c**), the resulting corrected skeleton counts do not match the expectation of all width levels having the same 25% probability. This discrepancy is caused by skeleton segment encroachment from higher-level branches through the branch connection boundary into lower-level branches (**Fig. S9**). Encroachment is problematic because this extension increases the length of the skeleton segment of a given object branch (**Fig. S8d**). Additionally, the width measurements contributed by the pixels within the encroachment area do not accurately represent the width of either of the branching levels. For this reason, SST is used to remove skeleton pixels between branch levels by trimming the skeleton segment pixels near the junction points.

A five-level hierarchical branched structure, with additional structural variability and complexity (e.g., curved branches, orientation distribution, branch length distributions, overlapping fibrils) was analyzed by FACT (**Fig. 6**). The width of the five branch levels was set at 4, 9, 18, 36, and 62 pixels (**Fig. 6**). FACT measured distinct width distributions for each level, the mean and standard deviation of which were 4.5(0.6), 9.2(1.0), 17.1(1.3), 35.6(0.7), 62.3(0.8) pixels, respectively. (**Fig. 6b**). The orientation correction had minimal influence on width 4.5(0.6), 9.2(1.0), 17.2(1.3), 35.6(0.7), 62.3(0.8) pixels, as all the branches were off-axis to the image coordinate axis (**Fig. 6c**). The broader width distribution for each branching level as compared to the four-level rectangular branched structure in **Fig. 5**, was a result of variability in widths of the lines drawn, the curved fibril profile, edge effects, and higher fraction of fibrils with off-axis orientation.

In summary, FACT width distribution measurements performed best for objects with aspect ratios greater than 10 and pixel densities greater than 5 across their width. As fibril orientation becomes more randomized with respect to the image coordinate axis, the effects of orientation correction on pixel counts and width statistics are diminished. For the hierarchical branched structures, all branching levels were sufficiently represented via skeleton segments, and each level contributed pixels to the width histogram proportional to its total length and aspect ratio. However, estimating branch length using pixel counts is not recommended, as discontinuities (e.g., particle edge effects, branch encroachments, and overlapping branches) influence the formation of skeleton junction points. Thus, the length of a skeleton segment does not represent the length of a given fibril element. The implementation of skeleton refinement (SST and SSF) helps to remove spurious skeleton segments that form from discontinuities and trims unwanted skeleton pixels in the vicinity of junction points.

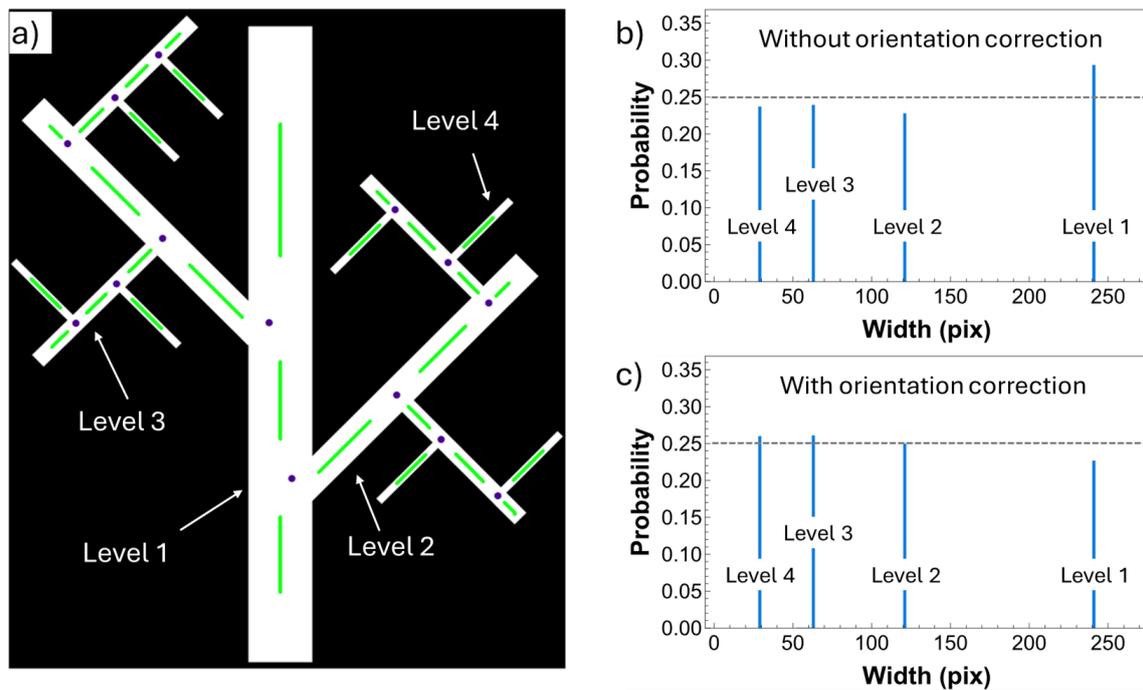

**Fig. 5** Simulated four-level hierarchical branched structure using rectangular segments (image size of 2107 x 2468 pixels). **a** FACT analyzed image with skeleton segment (green lines) and junction (purple dots) overlay. **b** FACT width assessment without orientation correction. **c** FACT width assessment with orientation correction. The histogram bin width is 2 pixels. SST = 50% and SSF = 10°

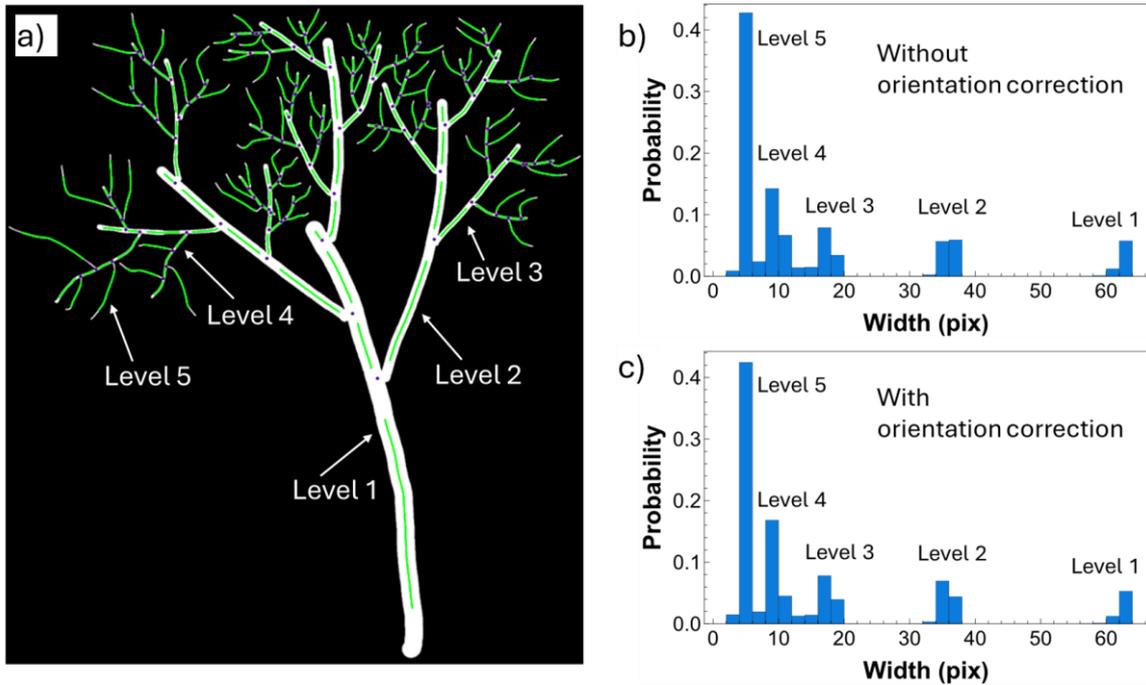

**Fig. 6** Simulated five-level hierarchical branched structure using curved segments (image size of 2016 x 2319 pixels). **a** FACT analyzed image with skeleton segment (green lines) and junction (purple points) overlay. **b** FACT width assessment without orientation correction. **c** FACT width assessment with orientation correction. The histogram bin width is 2 pixels. SST = 30% and SSF = 10°

## Validation with Wire Micrographs

The FACT approach was used to measure the width of small-gauge fixed-diameter wires (37 and 39 ga). The wires were placed on an optical imaging calibration standard (1 mm) and imaged with an I4-infinity optical microscope under backlighting. Image resolution was 2.03 μm/pixel. The Line tool in ImageJ (FIJI) was used to manually measure the width of the raw optical images, and measurements were taken at 10 locations along each wire, perpendicular to the longitudinal axis of the wire. The manual measurements yielded an average and standard deviation for the 39 ga and 37 ga wires of 92.4 μm (1.6 μm) and 124.0 μm (0.9 μm), respectively.

For the FACT analysis of these wires, images were segmented by grayscale thresholding using ImageJ. FACT was used for morphological thinning, skeleton refining, and width measurement of the wire segments. To refine the skeleton segments, an SST fraction of 0% and an SSF angle of 1 degree were used on these images. As shown in **Fig. 7,** the wire segments are identified with white shading, with the skeleton segments overlayed as green lines. The FACT width measurements were based on 1044 and 1022 points along the skeleton segments, resulting in a mean width of 89.4 (1.9) μm and 118.2 (1.5) μm for the 39 ga and 37 ga wires, respectively. The slight discrepancy between manual and FACT measurements was mainly attributed to the noise floor of these measurements, as the wire images had a pixel size (2.03 μm/pixel), which is about 2% of the total object's width. Nevertheless, the tight width distribution (**Fig. 7c and d**) indicates measurement consistency.

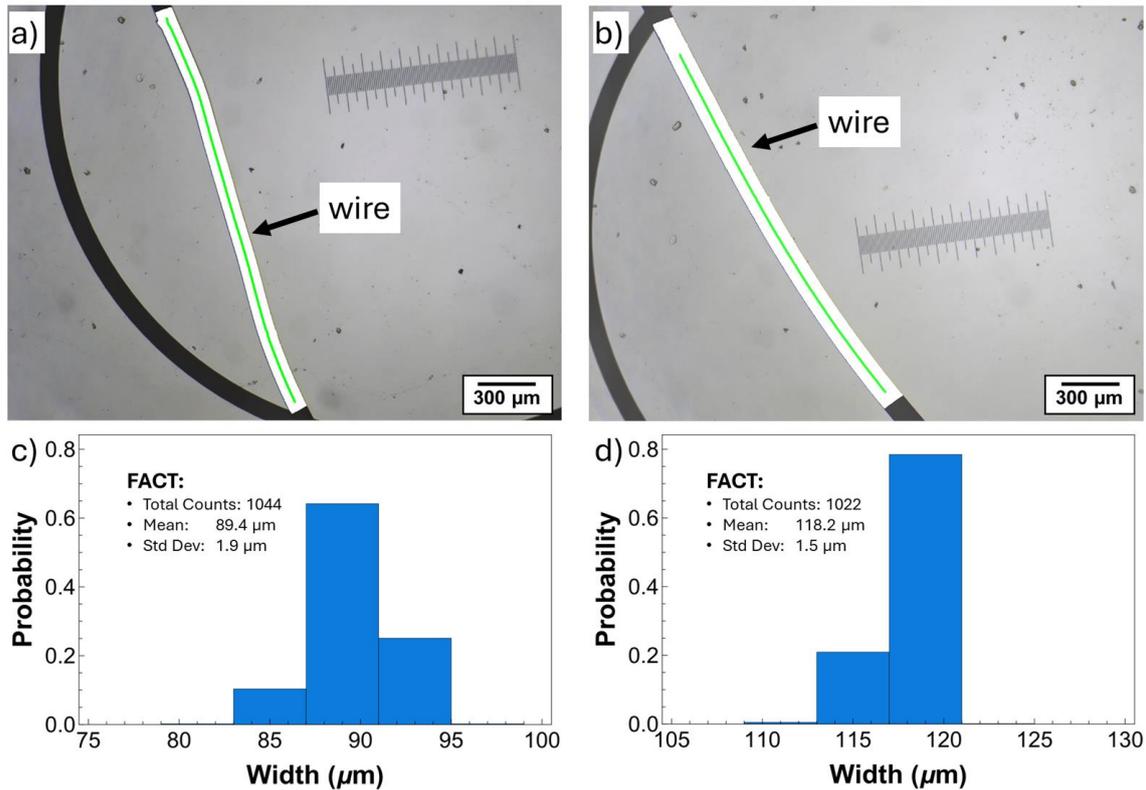

**Fig. 7** FACT analyzed optical images of small gauged wires: **a, c** 37 gauge wire, and **b, d** 39 gauge wire. The wire area analyzed is shaded white, with the skeleton segment overlayed as green lines. The histogram bin width is 4 μm. SST = 0% and SSF = 1°. Image pixel resolution: 2.03 μm/pixel

**Low-Level Branching CNFs**

To assess the effectiveness of FACT in measuring the width distributions of isolated, low-level branched CNFs, the FACT approach was applied to five NegC-SEM images captured in the study by Beaumont *et al*. (Beaumont et al. 2021). The resulting images had sufficient intensity contrast between CNFs (dark) and the substrate (medium grey), a low distribution density of CNFs, and a low level of CNF entanglement and branching. Such images are advantageous for FACT analysis (**Fig. 8a**), allowing for an unambiguous segmentation of the fibrils using Weka. FACT successfully identified and segmented the CNF objects, as demonstrated by the direct overlap of the segmentation image (purple overlay) shown in **Fig. 8b**. Careful inspection shows neighboring substrate pixels were captured in the fibril segmentation, resulting in a 1-to-2-pixel dilation of the fibrils and roughened fibril edge surface. This effect will increase the width measurement and promote the formation of erroneous skeleton segments and skeleton junctions. These issues could be minimized by either imaging at higher magnification to increase pixel density across the fibril width or by improving segmentation using the U-Net algorithm. Since there were insufficient images to train the U-Net, the current analysis utilized Weka segmentation, followed by the application of a Gaussian filter with a kernel size of 2 to remove some edge roughness. The morphological thinning and skeleton refining (SST = 10% and SSF = 20 degrees) resulted in skeleton segments that predominately traced the centerlines of the CNF structures, as shown via the green skeleton overlay (**Fig. 8c**). FACT correctly did not segment the white objects of unknown origin that were not CNFs.

The high aspect ratio of the fibrils resulted in skeleton segments encompassing much of the fibril length. The SST and SSF skeleton refining techniques were effective in removing many of the unwanted skeleton segments. The skeleton junctions (purple dots) were created at fibril branching or overlapping, but also created by other features (e.g., roughened fibril edges, kinks in fibril profile, etc.). Note that extra junction points complicate associations to the density of branching or the fibril network structure, as well as relating skeleton segments to the number and/or length of fibrils (**Fig. S7**) or in the case of network structures fibril element lengths. For example, examining the unbranched fibril labeled "*", it consists of 5 junction points, which are not associated with branching or fibril crossover points, and the fibril is effectively cut into 6 skeleton segments. The ramifications of this are that the image analysis of this fibril could be misrepresented. This single, isolated fibril, approximately 300 nm in length, could be incorrectly interpreted as branching or part of a network structure, and is composed of 6 fibril elements with lengths ranging from 20 to 200 nm. To avoid such errors in image analysis, the authors recommend using the skeleton segment points to measure the widths of the entire fibril structure and using extreme caution when considering any branching and/or fibril element length analysis.

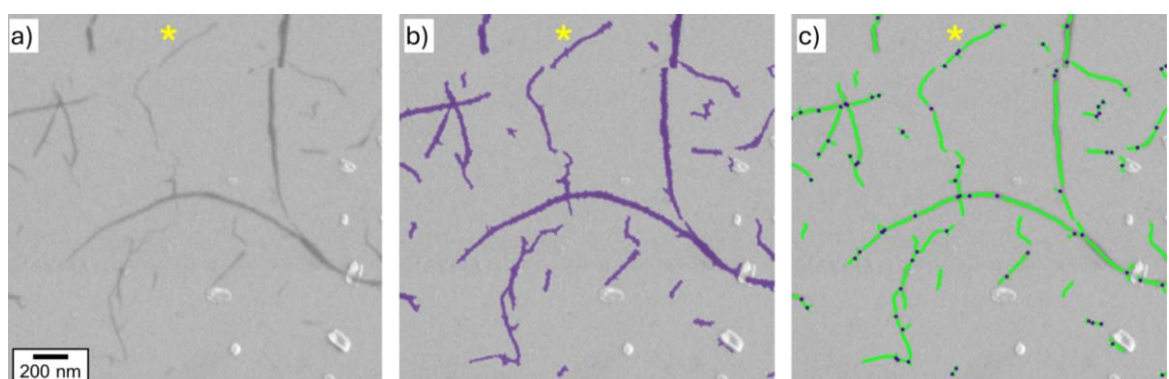

**Fig. 8** FACT analysis of NegC-SEM image of low-level branched CNFs from the study by Beaumont *et al.* (Beaumont et al. 2021). **a** As-received NegC-SEM image, digitally zoomed-in region, showing adequate contrast between CNFs and substrate background. **b** FACT analyzed image segmentation (purple) overlaid on NegC-SEM image, showing good object identification. **c** FACT skeleton segmentation (green line overlay) and junction points (purple dots) of fibrils (both are dilated for ease of view). SST = 10% and SSF = 20°. Image pixel resolution: 5.43 nm/pixel

Manual measurements of these five images were completed separately from the Beaumont *et al.* study (Beaumont et al. 2021) by one of their team members. Overall, the FACT and manual image analyses were in reasonable agreement, with similar width distributions, means, and standard deviations, where the differences in means were within the image pixel resolution. At lower magnification (**Fig. 9**, having a 5.43nm/pixel resolution), the mean width and standard deviation were: FACT 22.1 (8.9) nm, manual 22.7 (8.2) nm. The counts for FACT at 0-10 bin width would result from fibrils having a width of 1 pixel. At higher magnification (**Fig. 10**, having a 1.79 nm/pixel resolution), the mean width and standard deviation were: FACT 16.9 (5.2) nm, manual 17.1 (2.5) nm. The higher magnification image resulted in a narrower width distribution, lower mean width, and lower standard deviation. This result can be attributed to the more uniform size of CNFs captured within the image, improved segmentation, and higher image pixel resolution across the fibril width.

The "correctness" of image analysis relies on the premise that the images analyzed provide a reasonable representation of the CNF size distribution. For example, two additional images were analyzed at the lower magnification configuration, which included larger fibrils. The image analysis of all three images, **Fig. S10**, resulted in a broadening of the width distribution and an increase in the mean width and standard deviation: FACT 24.8 (15.2) nm, manual 23.1 (15.2) nm. All of which indicates the inclusion of wider fibrils in the analysis. A typical strategy for capturing a reasonable representation of CNF sizes is to analyze multiple images across a range of magnifications (Moon et al. 2023; Moon et al. 2025).

Despite notable differences between FACT and manual approaches, the resulting width analyses were similar. In the manual approach, each fibril is measured once in the mid-section using ImageJ, resulting in significantly lower counts. In contrast, the FACT approach obtains thousands of counts from the total collection of pixels that comprise the entire skeleton image. Thus, any given fibril will be measured multiple times along its length. The resulting data is not relatable to a specific fibril but to the "fibril collective" within the image.

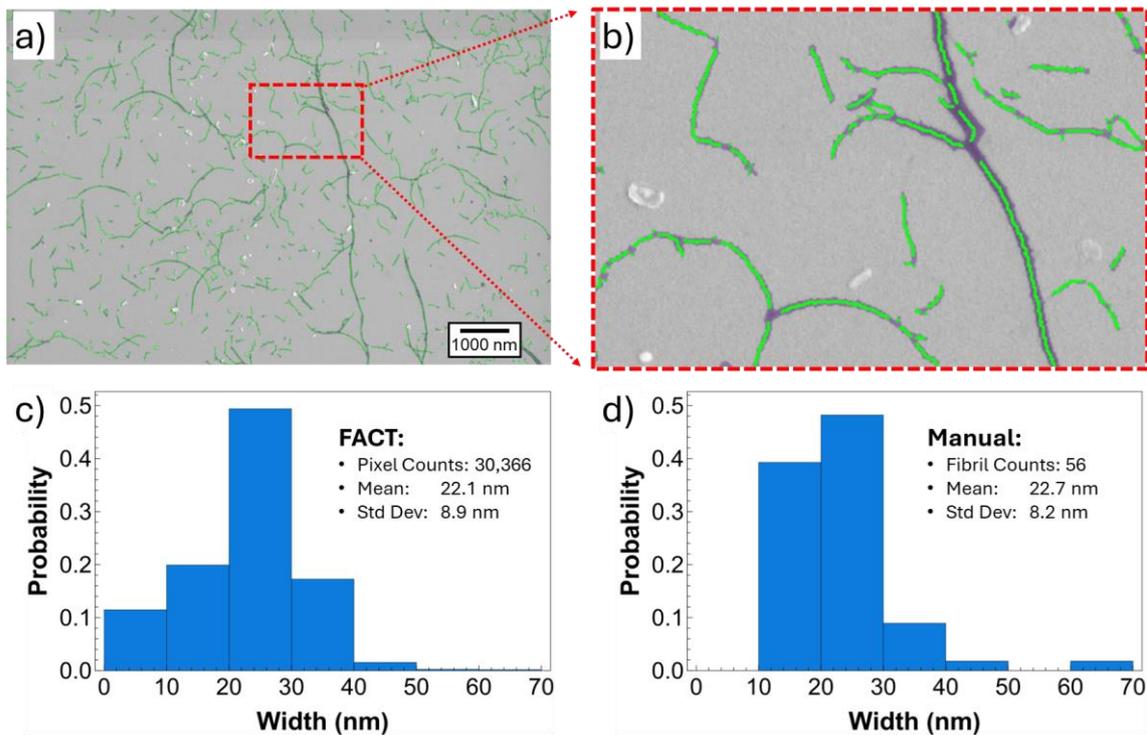

**Fig. 9** Image analysis of lower magnification NegC-SEM images of low-level branched CNFs from the study by Beaumont *et al.* (Beaumont et al. 2021). **a** FACT analyzed image with the FACT segmentation (pale purple) and the skeleton (green lines) overlays. The red rectangular overlay shows the location of the digital zoomed regions for part b. **b** Digitally zoomed-in region showing good agreement of FACT object identification, segmentation, and skeletonization of CNF fibril structure. **c** FACT width measurements for all pixels that make up the green skeleton overlay within part a. **d** Manual width measurement of individual fibrils. The histogram bin width is 10 nm. SST = 10% and SSF = 20°. Image pixel resolution: 5.43 nm/pixel

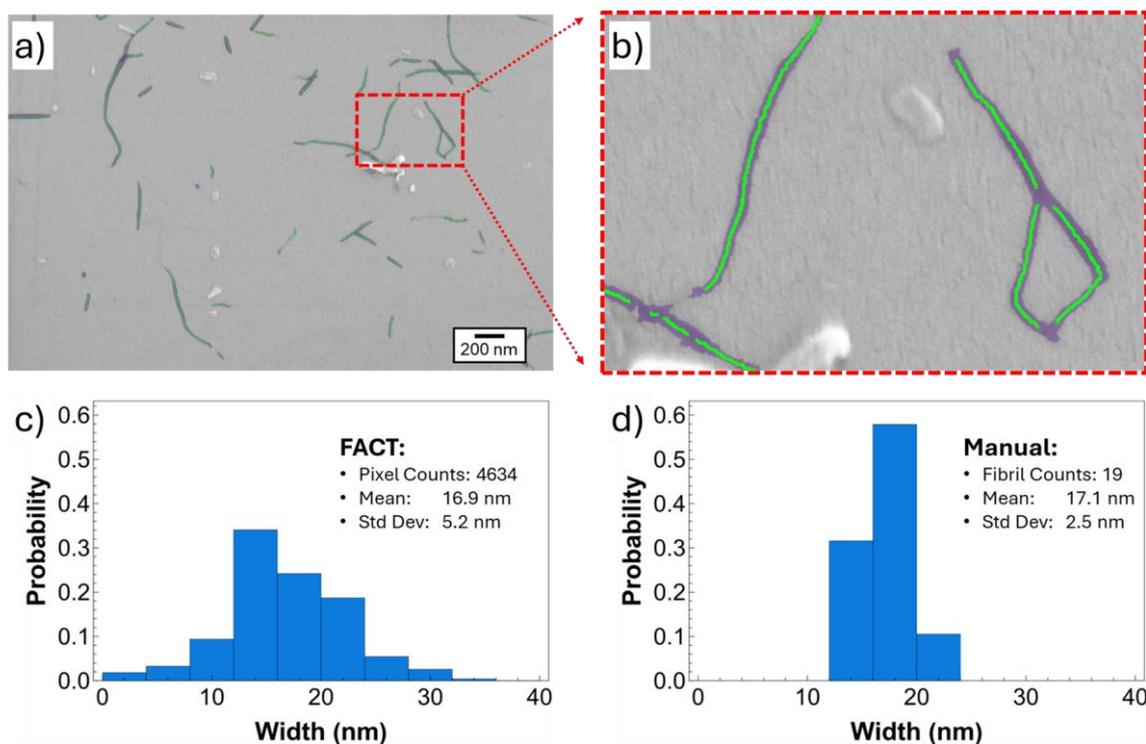

**Fig. 10** Image analysis of a higher magnification NegC-SEM image of low-level branched CNFs from the study by Beaumont *et al.* (Beaumont et al. 2021). **a** FACT analyzed image with the FACT segmentation (pale purple) and the green skeleton overlays. The red rectangular overlay shows the location of the digitally zoomed region for part b. **b** Digitally zoomed-in region showing good agreement of FACT object identification, segmentation (pale purple), and skeletonization (green line) overlays of CNF fibril structure. **c** FACT width measurements for all pixels that make up the green skeleton overlay within part a. **d** Manual width measurement of individual fibrils. The histogram bin width is 4 nm. SST = 10% and SSF = 20°. Image pixel resolution: 1.79 nm/pixel

**High-Level Branching CNFs**

To assess the effectiveness of measuring the width distributions of highly branched and networked CNF materials, the FACT approach was applied to NegC-SEM images captured in the study by Ringania et al. (2022). These complex fibril structures are typical of mechanically refined wood pulp fibers (**Fig.11**). The coarser CNFs and network structure required imaging at lower magnifications as compared to the CNF materials analyzed in the prior section. The U-Net segmentation and thinning operations result in an adequate skeleton segment trace of the centerline for most fibrils. To further reduce the effect of particle edge roughness, a Gaussian filter with a kernel size of 2 was applied to the resulting U-Net segmented images. For thin fibrils, the skeleton overlay showed good agreement with the CNF network (**Fig. 11c and d**), indicating that the subsequent width assessment will be accurate. However, in regions of extensive fibril bundling, branching, or overlapping, as shown in the center of **Fig. 11e**, there is some discrepancy between the green skeleton segments and the fibril network. This discrepancy is a result of several factors: the encroachment of the higher-level branch segments into the lower-level segments (described in **Fig. S9**), image segmentation issues in this area due to context ambiguity, and the thinning operator's sensitivity to sharp edges along the fibril boundary, which can result in extraneous branches (**Fig. 4**). Such deviations will result in errors in the width assessment of these regions. To reduce measurement errors, the refining steps SST and SSF are used to remove pixels from encroachment skeleton segments and remove extraneous branches, respectively. Optimizing the SST and SSF parameters to maintain a high fraction of the correct segments is an imperfect process, as some extraneous skeleton segments will inevitably remain, and if too much filtering is used, many correct segments could also be removed. A robust approach to quantifying the fraction of unwanted skeleton segments remains unclear.

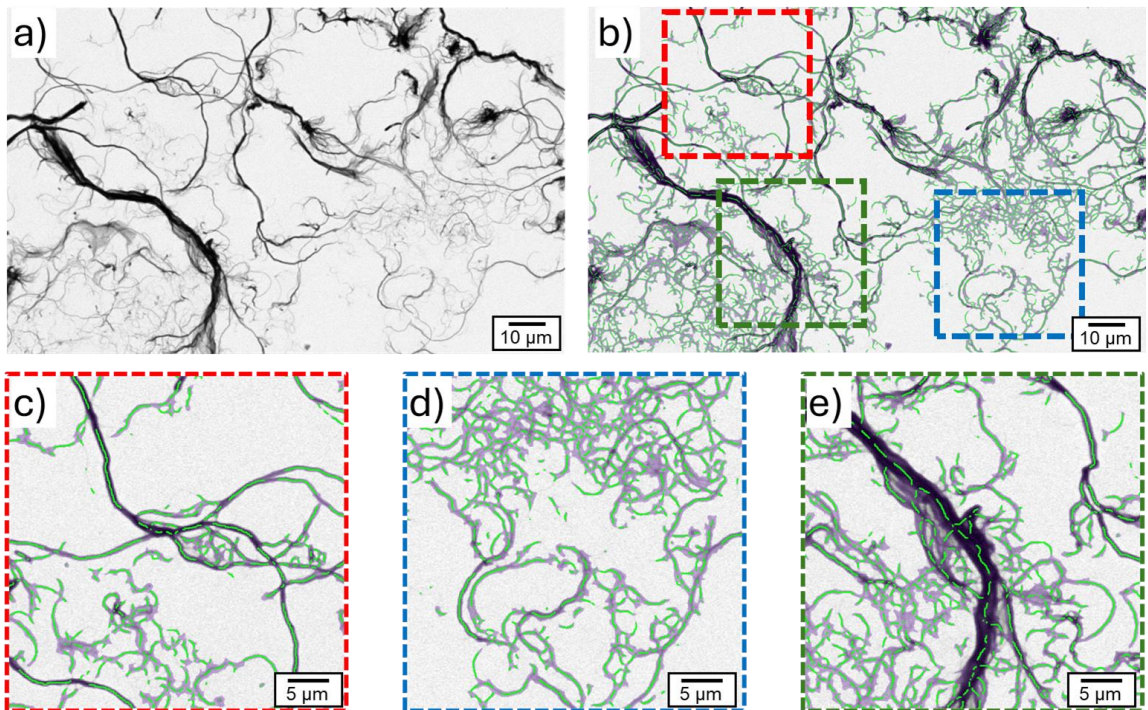

**Fig. 11** FACT analysis of NegC-SEM image of high-level branched CNFs from the study by Ringania (Ringania 2023).**a** Raw NegC-SEM image. **b** FACT analyzed image showing both segmentation (pale purple) and the skeleton segment (green lines) overlays. The three boxed overlays show the location of digitally zoomed regions of parts c, d, and e. **c** and **d** Skeleton segment overlays showing good agreement with the thinner fibril elements. **e** Skeleton segment overlay of thick fibril bundle and network region, showing a moderate level of discontinuity in representing the actual fibril structure. SST = 25% and SSF = 25°. Image pixel resolution: 80 nm/pixel

Coarser CNF fibril structures require imaging over a range of magnifications to analyze both the coarse and fine fibril structures, as demonstrated in **Fig. 12**. Three magnifications were used, with image pixel resolutions of 131, 38, and 22 nm/pixel. The width distribution was non-Gaussian, being skewed towards higher widths. With higher magnification, the width distributions shifted to lower values, as did their corresponding means and standard deviations (Low: 1234 (1259) nm, medium: 628 (534) nm, and high: 436 (348) nm). The shifting to narrower fibril widths with higher imaging magnification is considered to have resulted from three primary factors: 1) image positioning to a region with finer fibrils, 2) optimized image contrast, segmentation, and skeletonization of the finer fibril regions, improving the identification of faint fibrils, and 3) improved fidelity in fibril width measurement resulting from higher pixels density across the fibril width. Digital zoom would primarily improve factors 1 and 2, but not factor 3. At each magnification, there are regions where the green skeleton segments do not perfectly match the coarse fibril bundling, extensive branching, and overlapping fibrils. The skeleton segments are fragmented, resulting from encroachment of the higher-level skeleton into the lower-level segments and extraneous branches produced by sharp edges along the fibril boundary. The effects on fibril width measurement were reduced by using SST = 25% and SSF = 25° to trim the lengths of encroachments and remove most of the extraneous branches, respectively (**Fig. 12c**).

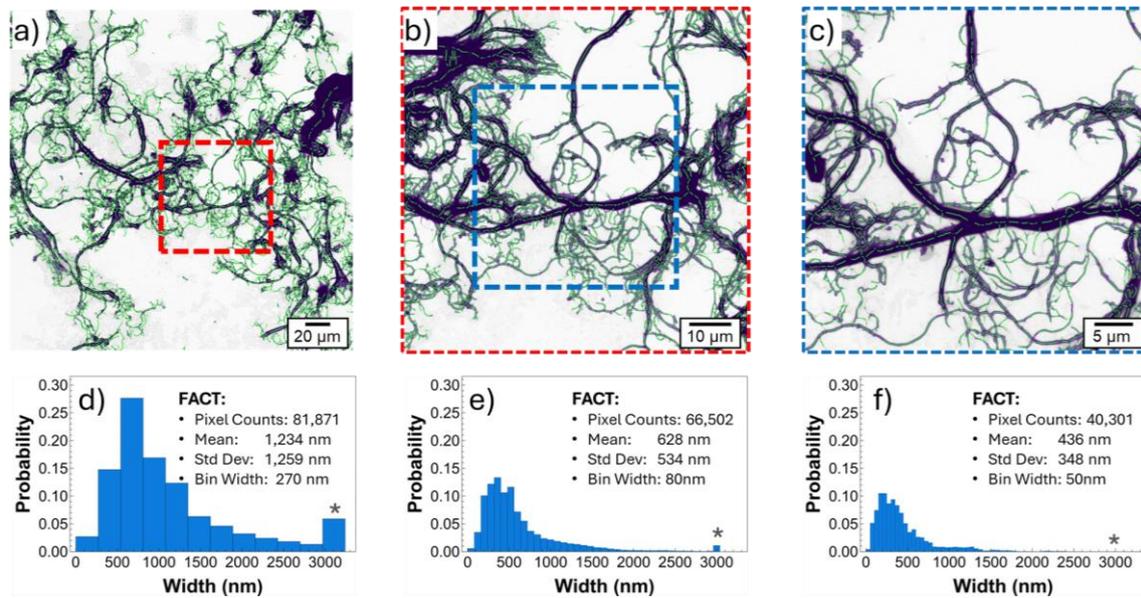

**Fig. 12** FACT analysis of NegC-SEM image of high-level branched CNFs from the study by Ringania *et al.* (Ringania et al. 2022) at three different magnifications. **a** 131 nm/pixel, **b** 38 nm/pixel, and **c** 22 nm/pixel. FACT analyzed images with binary segmentation (pale purple) and skeleton overlay (green lines), showing good agreement of the skeleton overlay with the CNF fibril structure. **d**, **e**, and **f** FACT fibril width distribution for the corresponding three magnifications, showing a shift to smaller fibril widths for higher magnification and image positioning to finer fibril regions. * Bin for all width measurements greater than 3000 nm. The histogram bin width is approximately 2 times the pixel resolution. SST = 25% and SSF = 25°

     The FACT width measurements were compared to the manual measurements reported by Ringania *et al.* (Ringania et al. 2022) for the three images shown in **Fig. S11**. The contrast between the two measurement approaches is evident when comparing **Fig. 13a** and **Fig. 13b**. FACT analysis is based on thousands of pixel counts that make up the entire skeleton segmentation. The resulting data is not relatable to a specific fibril, but rather, it is a function of the entire "fibril network" captured within the analyzed images. Thus, the resulting FACT measurement histogram is based on the number of pixels of a given width or width range (**Fig. 13c**). In contrast, the manual measurements were of individual fibrils or branching fibril elements with a single measurement using ImageJ, the location of which are shown as red line overlays in **Fig. 13b**. The resulting manual measurement histogram is based on the number of fibrils of a given width or width range (**Fig. 13d**). This notable differences between FACT and manual measurement approaches will affect the final width distribution histograms.

     The FACT width distribution shown in **Fig. 13c** spans from 22 to 5160 nm, having a broad peak ranging from 50 to 600 nm, and a mean and standard deviation of 510 (532) nm, based on a total of 104,891 pixel counts. The manual width distribution shown in **Fig. 13d** spans from 64 to 4226 nm, having a broad peak ranging from 100 to 450 nm, and a mean and standard deviation of 320 (310) nm, based on 978 fibril elements measured. The overlap in the

peak range (100 nm to 450 nm) indicates consistency between the two approaches in identifying the dominant fibril widths. However, there are notable differences in which manual measurements are shifted to narrower widths, while FACT is shifted to larger widths. For manual measurements, the greater number of fine fibrils within the images resulted in more counts at narrower widths. In contrast, for FACT, multiple measurements are obtained from each fibril. The number of skeleton counts is, in part, dependent on fibril lengths, and since wider fibrils are generally longer, more pixel counts will result. This discrepancy indicates that FACT is not entirely commensurate with the manual measurement approach.

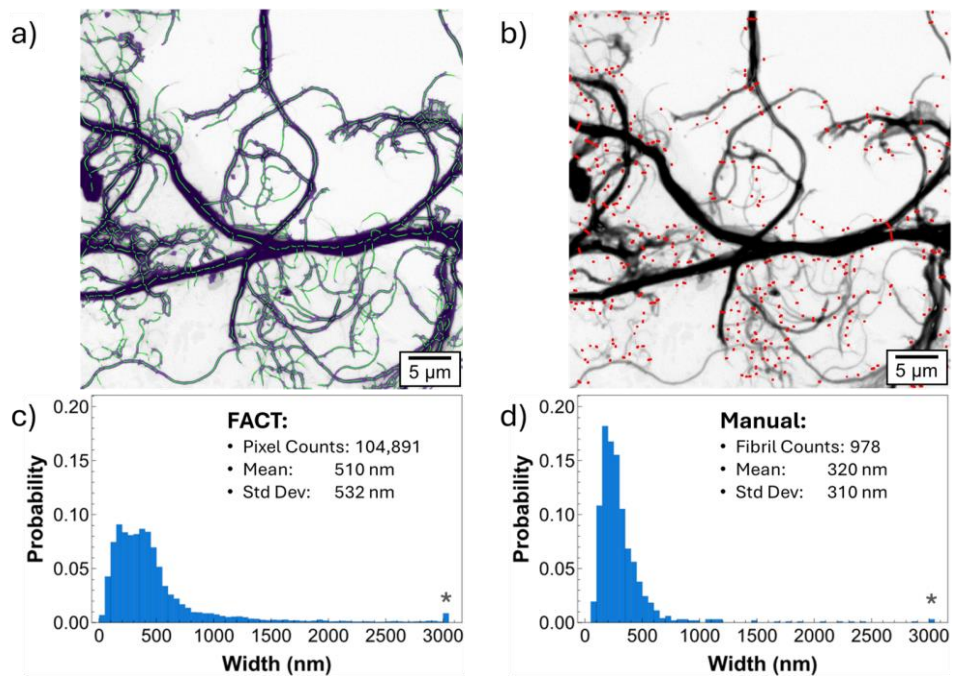

**Fig. 13** Comparison between FACT and manual fibril width measurement of high-level branched CNFs from the study by Ringania *et al.* (Ringania et al. 2022). **a** FACT analysis showing the skeletonization of the fibril network (green lines). ixels along these lines are used to measure width. **b** Manual analysis showing the fibril measurement locations (red lines are dilated for easier viewing). The width distributions based on image analysis of the three images in **Fig. S11** are shown for **c** FACT analysis and **d** manual analysis. * Bin for all width measurements greater than 3000 nm. Bin width of 50 nm. SST = 25% and SSF = 25°. Image pixel resolution: 22.0 nm/pixel

There are notable differences between FACT and manual measurement approaches; however, it is unclear which approach is the most effective for assessing the fibril width distribution of highly branched and networked CNF materials. Manual measurements give a single measurement for fibrils and branched fibrils. In general, fibril widths do not change much along their lengths unless there is branching. Thus, typically, a single width measurement could adequately describe a non-branching fibril. However, for highly branched and networked CNF materials, it can be challenging to confirm that a given fibril or branched fibril has only been measured once.

Additionally, analyst bias and fatigue will also skew measurement results. In contrast, for any given fibril measured with FACT, the number of skeleton pixel counts is proportional to its length and aspect ratio and inversely proportional to the number of junction points within. Fibrils with higher aspect ratios and length will proportionally contribute more skeleton counts to the statistical analysis of the population. This dependency indicates that the width count distribution measured with FACT is based on the proportion of an "effective fibril length" across all width ranges of the entire fibril network. This dependency on fibril length and aspect ratio differs from manual width distribution, which is based on the proportion of fibril numbers across all width ranges. Because of these differences in width measurement for FACT, the results require careful evaluation, as they may provide additional insights into the CNF particle and network morphology that are not feasible with manual methods.

**Network Assessment**

Relating FACT measurements (skeleton segments, skeleton junction points, and individual pixels) to specific features, such as fibril length, branching density, and network density, can be problematic. Skeleton junction point formation dictates the viability of such assessments. The issue is that skeleton junction formation is highly variable and dependent on the quality of image segmentation, skeletonization, skeleton refining, and the complexity of the fibril and network structures. In cases where the vast majority of junction points result from fibril intersections or branching (**Fig. 3**), it may be possible to correlate the number of junction points with the density of the fibril intersection network or the level of fibril branching. Likewise, if SST was set to 0, then the skeleton segments could be used to estimate fibril lengths or network segment lengths. However, as the proportion of junction points resulting from edge defects or segmentation imperfections increases, the relevance of the junction-point analysis decreases. For the low-branching CNF material, there were sufficient defects in the segmentation that the vast majority of junction points were from defects as opposed to fibril branching or intersections (**Fig. 8**). This was worse for the high-level branched CNF material. Currently, FACT lacks a practical approach for identifying and removing erroneous junction points.

**FACT Analysis Time**

A complete FACT analysis of a new image data set can take between 4 to 10 hours, depending on several factors, such as the number of images, image size, number of fibers per image, image segmentation approach (e.g., gray value threshold, Weka, or U-Net CNN), and the user's computer specifications. Analysis time can be categorized into several subtasks that follow the FACT flow chart in **Fig. 2**. The initial pre-analysis of incoming raw images takes ~ 1 min/image, in which FACT obtains pixel size using the scale marker within the image, trims the image to remove any boundary or banners, and saves the edited image as a new image file. Several edited gray images are segmented using Weka, an ImageJ plugin, which can take ~2 hours for an initial round of 4 images. If the user has only a small number of images to analyze and determines that the Weka segmentation results were adequate, they can continue to use FACT for width statistical assessment. However, if the user has a larger data set and determines that the segmentations could benefit from training a convolutional neural network, then the Weka segmentation can be used as the ground truth images to train a U-Net CNN. The time it takes to train a U-Net CNN depends on computer specifications and the number of ground truth images. This part could take between 2 to 8 hours. Finally, the FACT width analysis is fast, taking less than 2 mins per image, which includes the computing of various measurements for each skeleton segment and junction points (e.g., counts, point coordinates, distance transform values, centroid, and index), producing image overlays (e.g., segmentation, skeleton segment, junction points), and exporting the width data.

Although a complete FACT analysis can take a reasonable amount of time (e.g., 4 to 10 h), if the U-Net CNN is already trained, FACT can be used without additional training, significantly reducing analysis time to less than 5 min per image. A pre-trained CNN can be used if the incoming images have features, contrast, and noise levels similar to the original training images. Out of an abundance of caution, it is recommended to complete a test run using a pre-trained network to verify that FACT performs the segmentation and skeletonization adequately. Otherwise, it is possible to continue training the U-Net CNN with additional ground truth image files from the new dataset, but training with images of similar contrast, intensity, and noise is recommended.

## Conclusion

A semi-automated image analysis framework, Fibril Analysis for Cellulose Technology (FACT), was developed to rapidly and reliably measure the width distribution of fibril elements that make up cellulose nanofibril (CNF) particles. The high-contrast images from NegC-SEM enabled the capturing of the entire CNF hierarchical branching structure, spanning length scales from the micron-sized CNF object down to nano-sized fibril features. The FACT framework was validated using idealized geometries, simple wire micrographs, and hierarchical branched CNF structures. FACT width distribution measurements are most effective for fibrils with aspect ratios greater than 10 and a pixel density of at least 5 pixels across the fibril width. The number of skeleton pixel counts an object contributes is proportional to the object length and aspect ratio and inversely proportional to the number of junction points within the object. However, using these counts to estimate fibril branch length is not recommended due to the inconsistent nature of how discontinuities affect skeleton segments from junction points and branch encroachments.

FACT was successfully applied to two contrasting CNF morphologies (i.e., low and high branching), demonstrating effective segmentation, skeletonization, and measurement of fibril width. The FACT width measurement results were mainly in agreement with the manual measurements. Variations between the FACT and manual approaches were attributed to differences in the number of width measurement counts per fibril. For the manual, a single measurement is made per fibril. In contrast, for FACT, multiple measurements are made along the length of each fibril. The advantages of FACT are that the entire CNF branching and network structure is measured, the bias in fibril selection and measurement is removed, and it can potentially provide a higher level of detail not achievable with manual measurements. Thus, FACT analysis appears to be more advantageous than manual analysis for CNF materials with high-level branching and network structures. A complete FACT analysis can take 4 to 10 hrs. If the U-Net CNN is already trained, FACT can be used without additional training, significantly reducing analysis time to less than 5 minutes per image. The FACT code is publicly available in Zenodo. [Add Zenodo web link when published]

# Acknowledgment

The authors recognize Dr Ishant Tiwari from Chemical and Biomolecular Engineering at Georgia Institute of Technology for optical micrographs used in Fig. 7. The authors also recognize Dr. Bruno Mattos from the Department of Bioproducts and Biosystems at Aalto University for providing NegC-SEM images of their low-level branched CNF material and for conducting the subsequent manual measurements. We also gratefully acknowledge Dr Nayomi Plaza-Rodriques (Forest Products Laboratory), Dr Nathan Bechle (Forest Products Laboratory), Dr Warren Batchler (Monash University), and Dr Linda Johnston (National Research Council Canada) for critical reading of the document.

**Author Contributions**

The study conception was developed by U.R., C.B., R.M., and S.B.. The study design was developed by R.M., C.B., and U.R.. Coding design was developed by C.B.. Code testing and verification studies were completed by C.B., R.M., U.R.. Results analysis were completed by R.M., C.B., and U.R.. NegC-SEM imaging of high-level branched CNF material and subsequent manual width measurements were completed by U.R.. The main manuscript text was written by R.M., C.B., and U.R.. All authors critically reviewed a draft manuscript(s), which resulted in revisions to the document. All authors have read and approved the final manuscript.

# Declarations

**Funding**

This work was supported by USDA Forest Service-Forest Products laboratory, and by both the Renewable Bioproducts Institute and the Brook Byers Institute for Sustainable Systems at Georgia Institute of Technology.

**Conflicts of interests**

The authors declare no conflict of interest.

**Availability of data and material**

The FACT code is publicly available in Zenodo. [Add Zenodo web link when published]

**Consent for publication**

The authors consent for publication by the journal.

**Supplementary Material**

# Semi-automated image analysis of Cellulose Nanofibrils using Machine learning segmentation and Morphological thinning

Carlos Baez[1], Udita Ringania[2], Saad Bhamla[2], Robert J. Moon[1]

*1 The Forest Products Laboratory, USDA Forest Service, Madison, WI 53726*
*2 Chemical and Biomolecular Engineering, Georgia Institute of Technology, Atlanta, GA 30332*

**Terminology:**

- **Skeleton segments**. These are the lines that are produced through morphological thinning.
- **Skeleton junction points**. These are the skeleton junction points that coincide with CNF crossover points or fibril branching points.
- **Skeleton artificial junction points**. These are junction points caused by edge effects or other artifacts.
- **Morphological operators:** mathematical operators for the processing of binary images
- **Rastering:** the digitization/pixelation of a graphical object.
- **Fibril:** extremely high aspect ratio object, typically with uniform diameter along its length.
- **Branching:** when a smaller diameter fibril separates (emerges) from a larger diameter fibril, these are still bonded/held together at the branching point
- **Network:** When fibril crosses over itself or another fibril.
- **Branching vs Network:** FACT cannot differentiate between the two, but it is possible to set up reasonable differentiating criteria. For example, a branching fibril width should decrease after the junction point, and the branch should come off the "trunk" at shallow angles.
- **CNF morphology:** size and shape of a CNF network

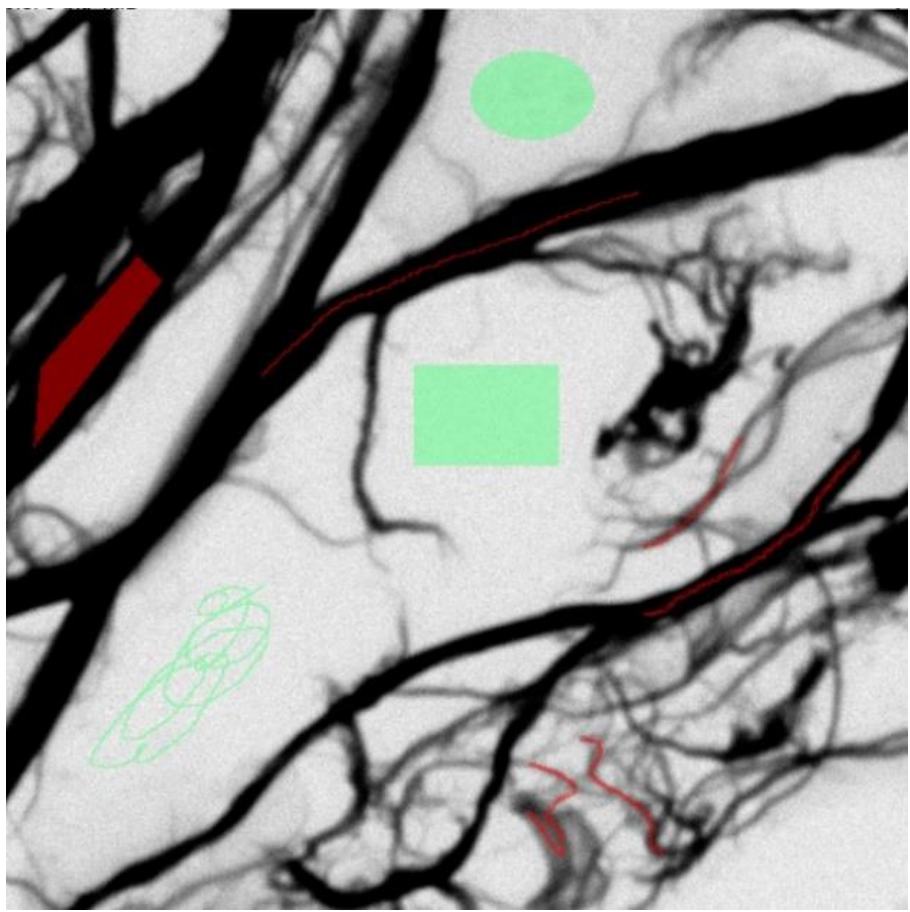

**Fig. S1** Raw NegC-SEM image showing the training using Weka in which regions of interest were defined with either shaded areas or lines for CNF fibrils (red) and background (green). Selecting more data can improve Weka's segmentation performance.

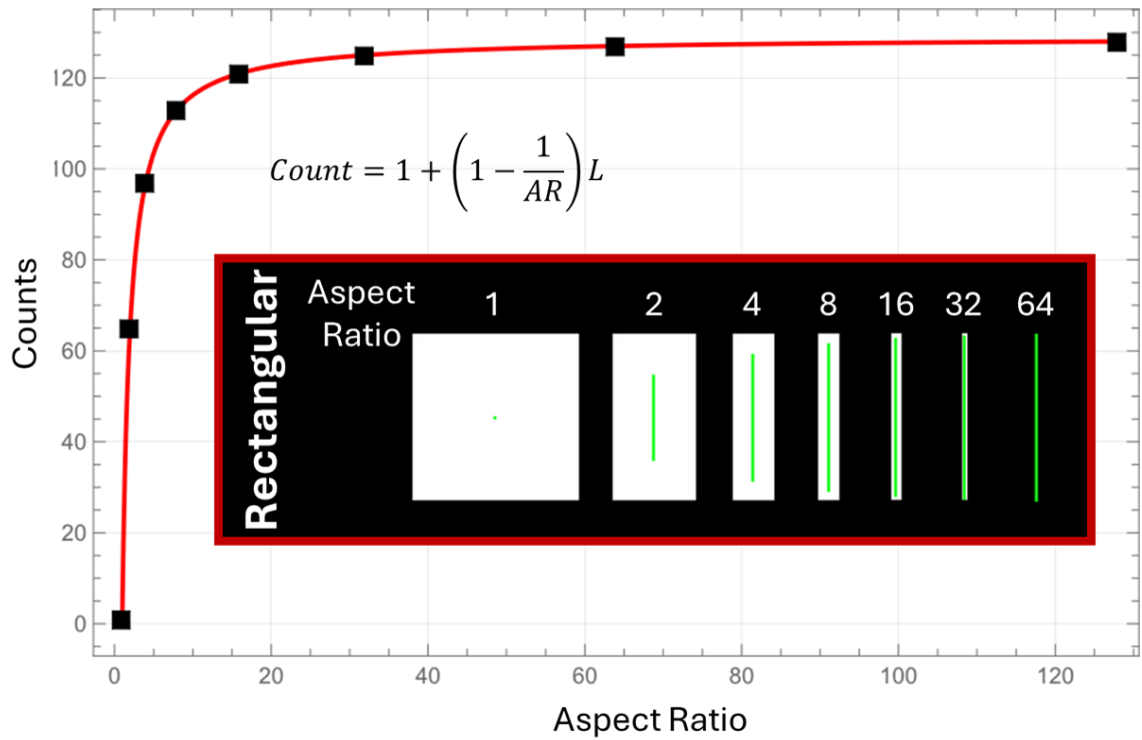

**Fig. S2** The morphological thinning approach to skeletonize an object removes pixels from all sides of an object. Thus, the number of pixels that make up the resulting skeleton segment depends on the object geometry. The insert gives a graphical representation of a rectangular object (pink color) with L = 128 pixels and its corresponding skeleton segment (green) (dilated so it is easier to see). Pixel removal around the peripheral of the object is limited by the smaller dimension (or side) and then converges into a central line. A square object with an aspect ratio of one will converge to a single pixel in the object's center. In contrast, a rectangular object with an aspect ratio of two will converge to a central line skeletal segment of 65 pixels. This plot is for objects with orientation on-axis with pixel rastering. SST = 0 and SSF = not applied

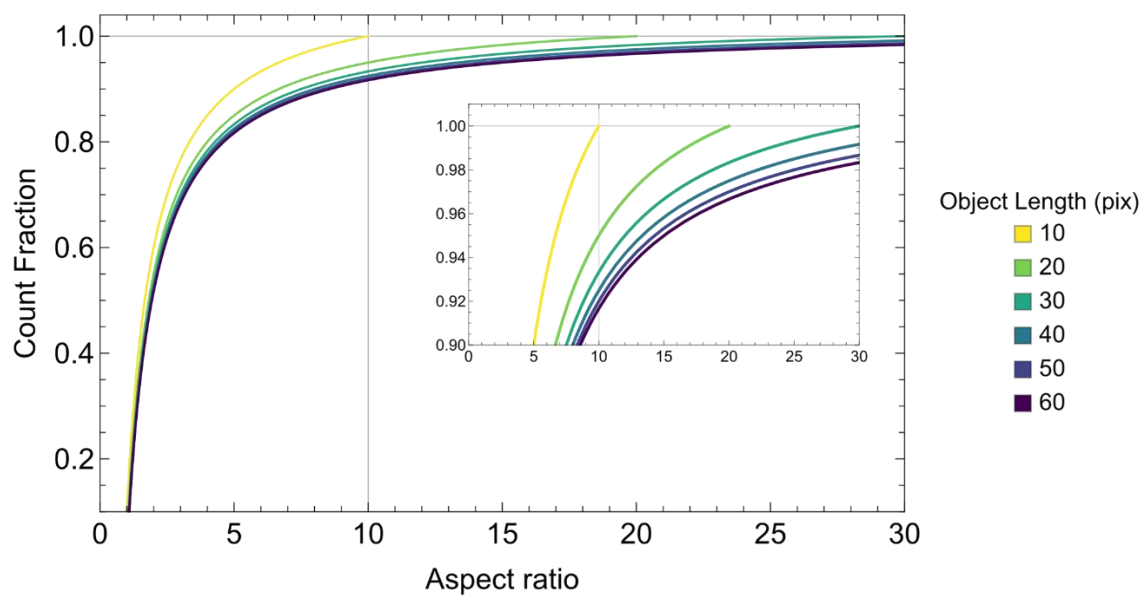

**Fig. S3** The effect of aspect ratio on the number of skeleton pixel counts as a fraction of the object length for rectangular objects of lengths (10, 20, 30, 40, 50, and 60 pixels). The skeleton pixel count of each object is normalized according to its length. Using FACT for width statistical analysis is most effective when the majority of objects within an image have aspect ratios greater than 10. When all objects have the same aspect ratio, shorter objects proportionally contribute more skeleton counts than longer objects. For CNFs, if we assume that all fibril segments have the same aspect ratio, shorter fibril segments will proportionally contribute slightly more skeleton counts than the longer segments. This effect is reduced when comparing fibril segments of higher aspect ratios (greater than 10). SST = 0 and SSF = not applied

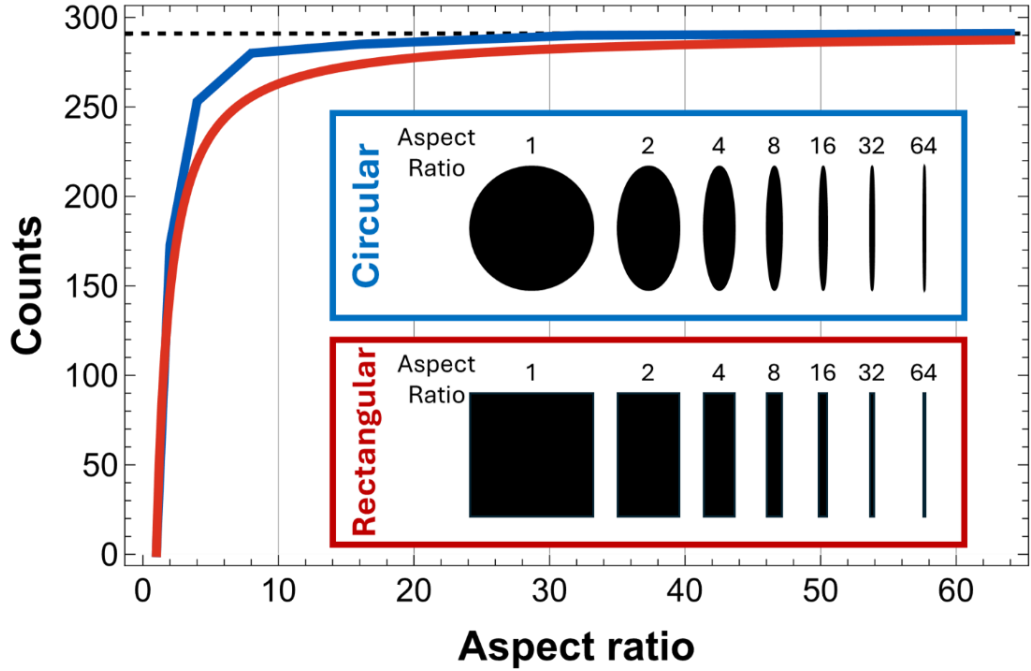

**Fig. S4** The effect of aspect ratio on the skeleton counts between objects with rounded edges (blue) and rectangular edges (red). In this example, the object's aspect ratio was doubled each time. Then, the resulting skeleton counts were measured. Here, the max length is 291 pixels. The rounded objects will generally contribute slightly greater counts at each aspect ratio level. Using FACT for width statistical analysis works best when most objects within an image have aspect ratios greater than 10 when analyzing objects with rounded edges. SST = 0 and SSF = not applied

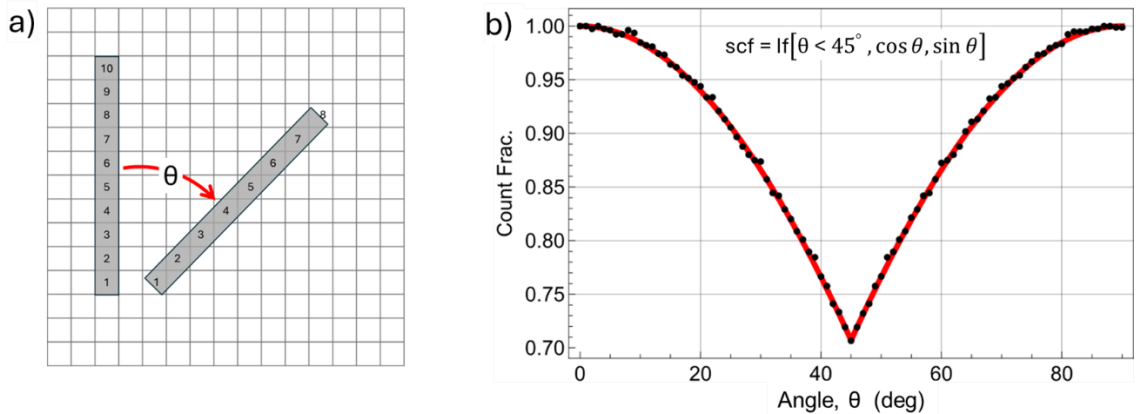

**Fig. S5** Effect of object orientation on skeleton segment pixel counts. **a** Schematic of 2 skeleton segments of equal length, the line orientated at 0 degrees with respect to the vertical axis consists of 10 pixels, in contrast to the line oriented at 45 degrees, which only consists of 8 pixels. **b** Normalized skeleton counts (black) of a rectangular object as a function of orientation with respect to the positive vertical axis. The fitting function (red) is used to compute a corrective factor. This corrective factor is used to resample the skeleton segment width values according to their orientation, accounting for this effect. SST = 0 and SSF = not applied

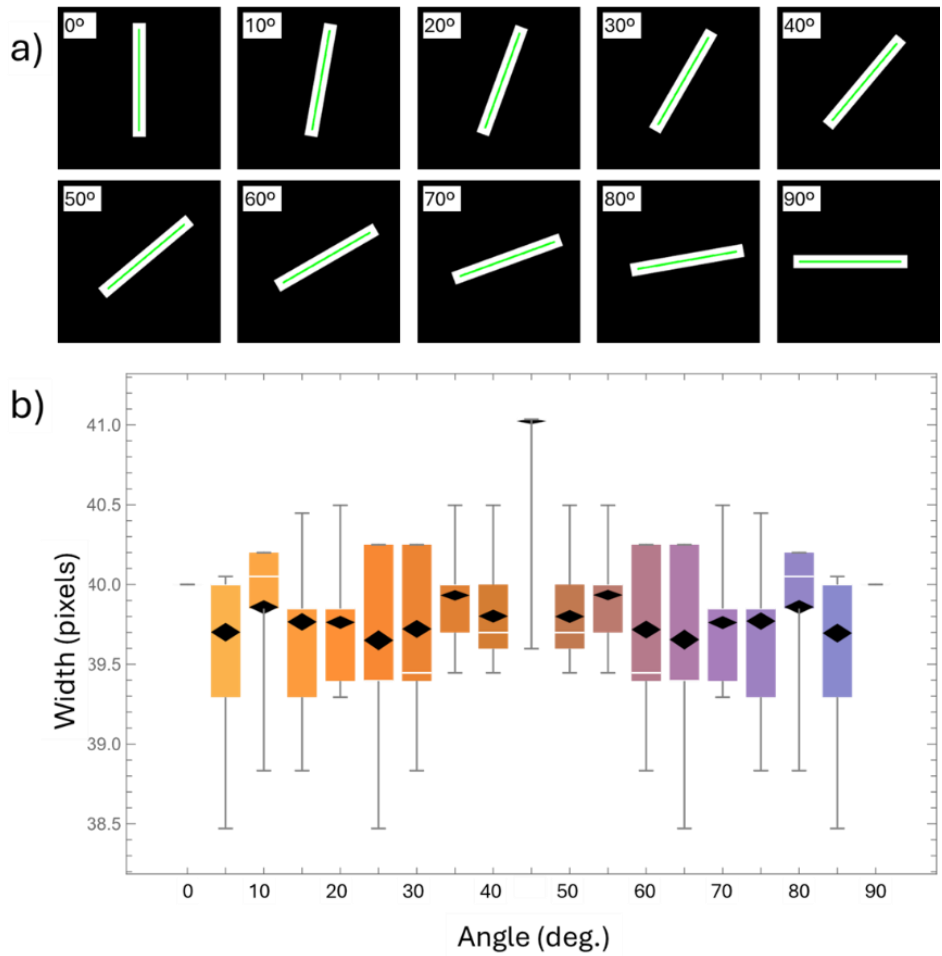

**Fig. S6** Effect of object orientation on width measurement. **a** Sequence of rectangular objects [length: 350 pixels, Width: 40 pixels] rotated from 0 to 90 degrees prior to rastering. The raster size is 500 by 500 pixels. **b** Box plots of the corresponding FACT width measurement for object orientation with respect to the horizontal axis. The diamond represents the mean. The variation in width measurement based on object orientation was small, ± 1.25 pixels. SST = 0 and SSF = not applied

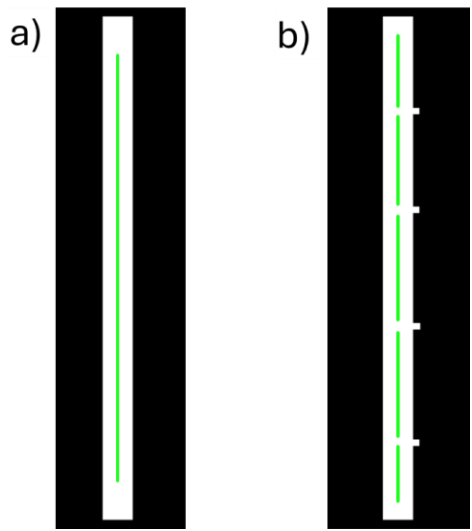

**Fig. S7** Effect of junction points on skelton segement, number, length and pixel counts. **a** Schematic fibril with zero junction points. FACT analysis calculated a single skelton segement consisted of a total of 1962 pixels. **b** Schematic fibril with 4 discontinues along its length. FACT alalysis calculated 4 junction points, 5 skelton segments having a combined 1950 pixel count. Box plots. SST = 10% and SSF = 10°

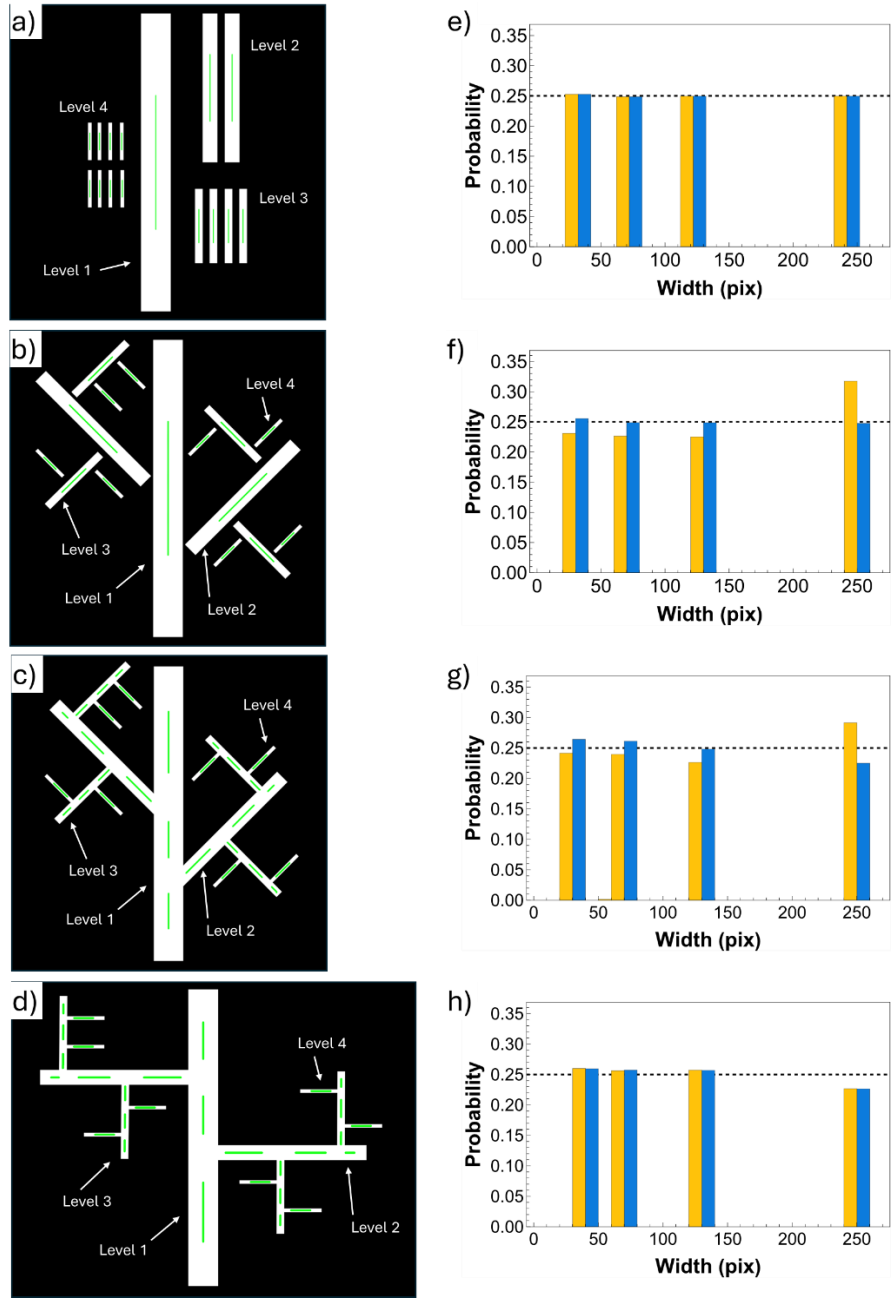

**Fig. S8** Effect of object orientation and branching on the width measurements counts of a 4-level simulated branched structure. Every branch level has an aspect ratio of 10, all levels have the same total length, and the width of the four branch levels were 30, 62, 120, and 240 pixels. **a** Isolated branches with all levels parallel to the vertical axis. **b** Isolated branches with levels 2, 3, and 4 orientated ± 45 degrees to the vertical axis. **c** Connected branches with levels 2, 3, and 4 orientated ± 45 degrees to the image vertical axis. **d** Connected branches with levels 1 and 3 oriented parallel to the vertical axis, and levels 2, and 4 orientated 90 degrees to the vertical axis. **e,f,g,h** FACT width assessment for each branching level for the corresponding branch arrangement in a, b, c, and d, respectively. The yellow bars represent the orientation uncorrected results, and the blue bars represent the orientation corrected result. The histogram bin width is 12 pixels, and the corrected data (blue) was shifted by 10 pixels for clarity. SST = 50% and SSF = 10°

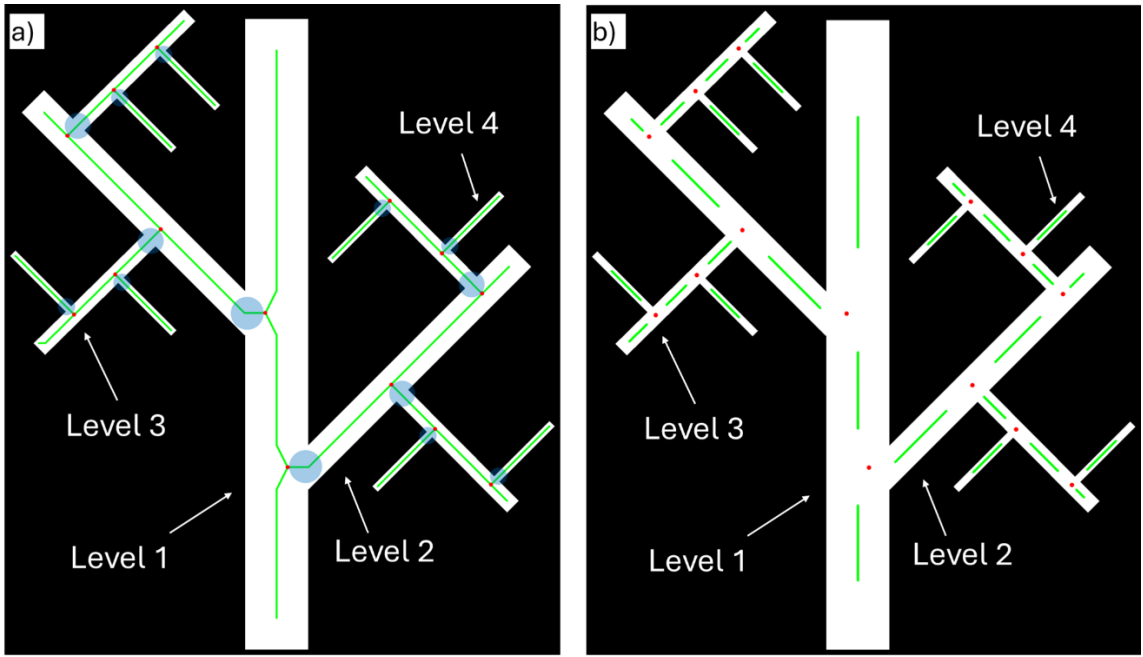

**Fig. S9** FACT skeleton refining of the 4-level simulated branched structure. **a** Skeleton segment trimming (SST) equal to 0. Highlighted here are the skeleton encroachment regions in blue-shaded circles (dilated so they are easier to see). Consider, for example, how the level 2 branch extends into the center line of the level 1 branch. Encroachment is problematic as this extension effectively increases the skeleton segment length of the higher-level branches. In addition, the width measurement of the pixels within these extensions is unwanted as it does not represent the actual width of the branch. **b** Skeleton refining with an SST = 50% removed unwanted pixels in this encroachment area by trimming the skeleton segments near the junction points.

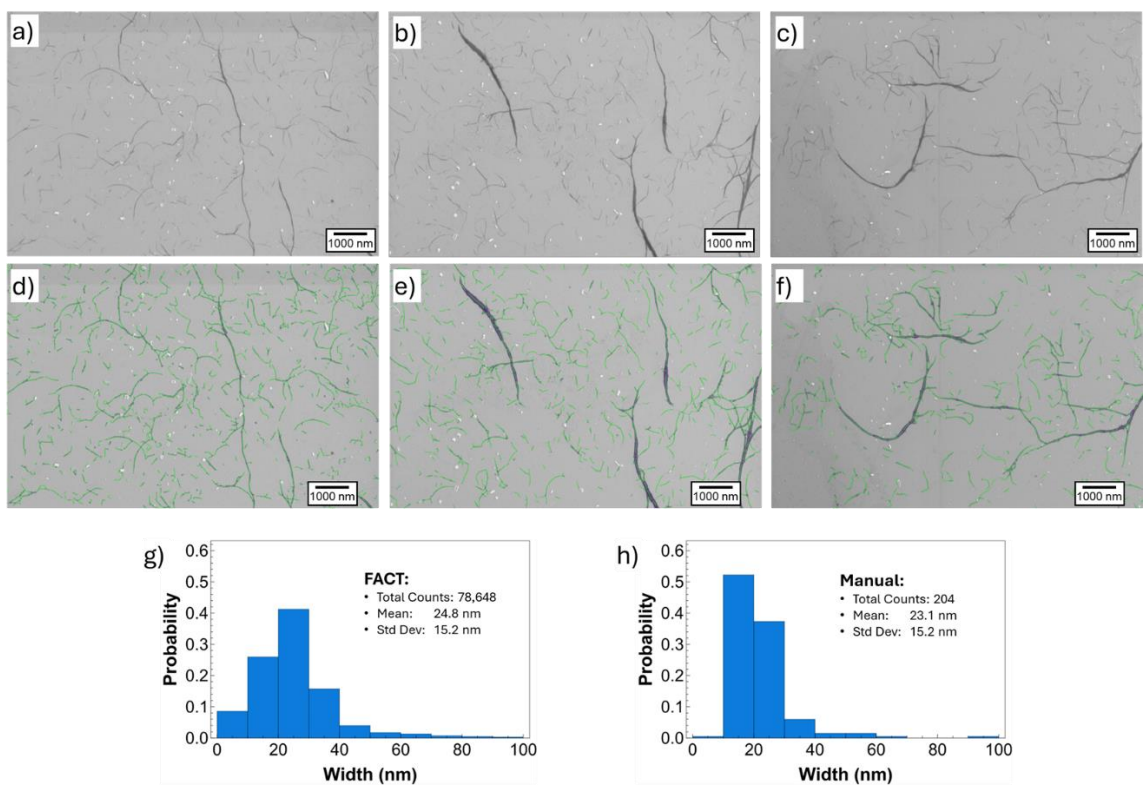

**Fig. S10** FACT analysis of NegC-SEM image of low-level branched CNFs from the study by Beaumont et al. (2021). **a, b, and c** As-received NegC-SEM images, showing good contrast between CNFs and substrate background. **d, e, and f** FACT analyzed images with FACT segmentation (pale purple) and the green refined skeleton overlay. **g** FACT width measurements for all pixels that comprise the green refined skeleton overlay, and **h** manual width measurement of individual fibrils. Bin width of 10 nm. SST =10% and SSF = 20°. Image pixel resolution: 5.43 nm/pixel

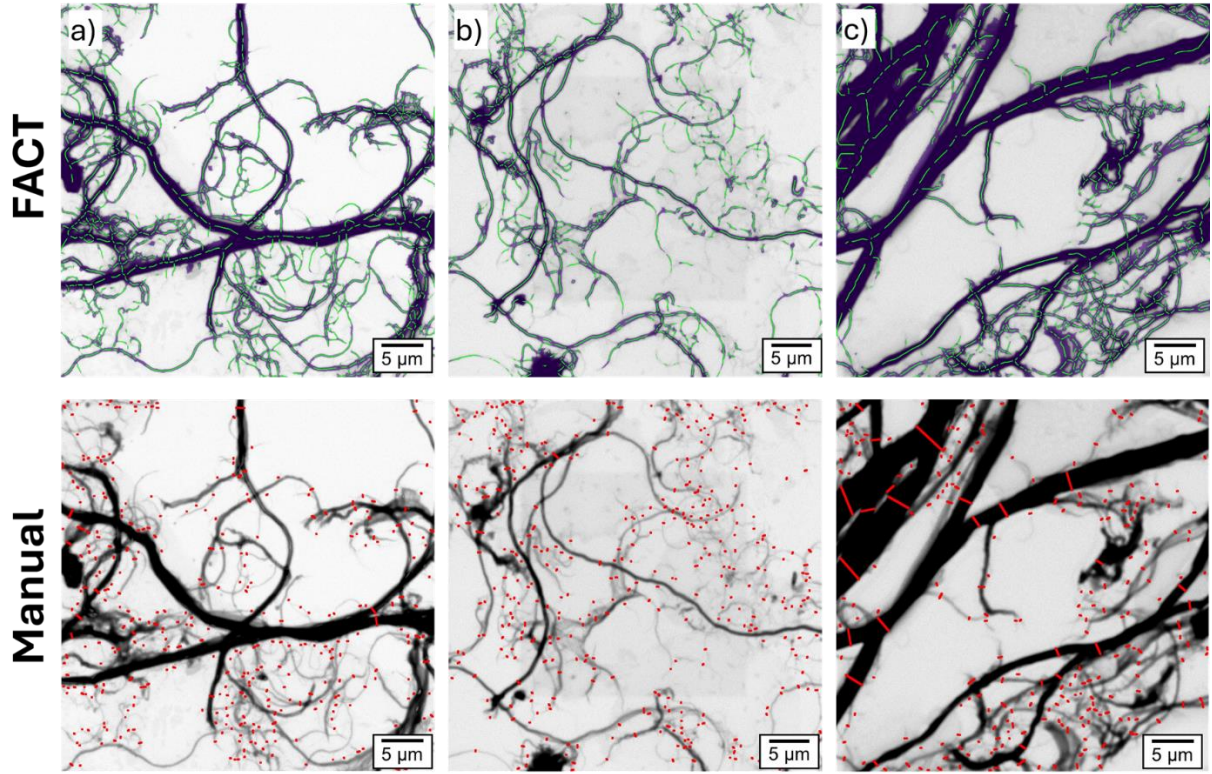

**Fig. S11** Comparison between manual and FACT fibril width measurement of high-level branched CNFs material from the study by Ringania et al. (2022). **a** image 1, **b** image 2, **c** image 3, all three images were analyzed with both manual and FACT image analysis. FACT analysis shows the refined skeleton of the fibril network (green lines). Pixels along these lines are used to measure width. SST =10% and SSF = 20°. Manual analysis shows the fibril measurement locations (red lines). The red lines were dilated for display purposes. Image pixel resolution: 22.0 nm/pixel